\documentclass[aps,pre,preprint,12pt]{revtex4}
\usepackage{epsfig}

\RequirePackage[T2A]{fontenc}
\normalsize

\begin{document}

\title{Steady mirror structures in a plasma with pressure anisotropy}
\author{E.A. Kuznetsov$^{(a,b,c)}$\/\thanks{%
e-mail:kuznetso@itp.ac.ru}, T. Passot $^{(d)}$, V.P. Ruban $^{(c)}$ and P.L.
Sulem $^{(d)}$}
\affiliation{{\small \textit{$^{(a)}$P.N. Lebedev Physical Institute RAS, 53 Leninsky
Ave., 119991 Moscow, Russia}}\\
{\small \textit{$^{(b)}$Space Research Institute RAS, 84/32 Profsoyuznaya
str., 117997, Moscow, Russia}}\\
{\small \textit{$^{(c)}$ L.D. Landau Institute for Theoretical Physics RAS,
2 Kosygin str., 119334 Moscow, Russia}}\\
{\small \textit{$^{(d)}$Universit\'e de Nice-Sophia Antipolis, CNRS,
Observatoire de la C\^ote d'Azur, PB 4229, 06304 Nice Cedex 4, France}}}
\date{}

\begin{abstract}
In the first part of this paper we present a review of our results concerning the weakly nonlinear regime of the mirror instability in the framework of an asymptotic model. This model belongs to the class of gradient type systems for which the free energy can only decrease in time. It reveals a behavior typical for subcritical bifurcations: below the mirror instability threshold, all localized stationary structures are unstable, while above threshold, the system displays a blow-up behavior. It is shown that taking the electrons into account (non-zero temperature) does not change the structure of the asymptotic model. For bi-Maxwellian distributions functions for both electrons and ions, the model predicts the formation of magnetic holes. The second part of this paper contains original results concerning two-dimensional steady mirror structures which can form in the saturated regime. 
In particular,  based on Grad-Shafranov-like equations, a gyrotropic plasma, where the
pressures in the static regime are only functions of the amplitude of the
local magnetic field, is shown to be amenable to a variational principle with
a free energy density given by the parallel tension. This approach is used
to demonstrate that small-amplitude static holes constructed slightly below
the mirror instability threshold identify with lump solitons of KPII
equation and turn out to be unstable. It is also shown that regularizing
effects such as finite Larmor radius corrections cannot be ignored in the
description of large-amplitude mirror structures. Using the gradient method, which is based on a variational principle for anisotropic MHD taking into account ion  finite Larmor radius effects, we found both one-dimensional magnetic structures in the form of stripes and two-dimensional bubbles when the magnetic field component transverse to the plane is increased. These structures realize minimum of the free energy.
\end{abstract}

\maketitle

\tableofcontents

\vspace{1cm}

\bigskip

\section{Introduction}

Magnetic structures in the form of holes or humps associated with maxima or
minima of plasma density and pressure are often encountered in planetary
magnetosheafs close to both the bow-shock and the magnetopause, and in the
solar wind (see e.g. \cite{Lucek01, Sper00, StevensKasper2007}) as well.
These structures are often viewed as ultra-low frequency (ULF) waves
resulting from the mirror instability (MI) \cite{VedenovSagdeev}, and, by
this reason, called mirror structures. This instability develops in a
collisionless plasma characterized by a
relatively large {$\beta $} (a few units) and a transverse (usually ionic) temperature {$%
T_{\perp }$} larger than the parallel one {$T_{\Vert }$}, such that the
condition for mirror instability {%
\begin{equation}
{T_{\perp }}/{T_{\Vert }}-1>{\beta _{\perp }^{-1}}  \label{thr}
\end{equation}%
is fulfilled. Here {$\beta _{\perp }=8\pi p_{\perp }/B^{2}$} (similarly, {$%
\beta _{\Vert }=8\pi p_{\Vert }/B^{2}$}), where {$p_{\perp }=n$}}${T_{\perp }%
}${and {$p_{\Vert }=n$}}${T_{\Vert }}$ are the perpendicular and parallel
plasma pressures respectively.

In the Earth magnetosheath, a typical depth of magnetic holes is about 20\%
of the mean magnetic field value and can sometimes achieve 50 \%. The
characteristic width of such structures is of the order of a few ion Larmor
radii, and they display an aspect ratio of about 7-10. In solar wind,
according to \cite{StevensKasper2007}, the size of holes may be very different,
varying from 10 up to 1000 ion gyroradii. In magnetosheath, holes and humps have 
comparable size. and amplitudes.
Humps are often observed near the magnetopause where {conditions (\ref{thr})
for development of the MI can be met under the effect of the plasma
compression.  Mirror structures are also observed when the plasma is
linearly stable \cite{EB96, Genot}, which may be viewed as the signature of
a bistability regime resulting from a subcritical bifurcation, whose
existence was interpreted on the basis of a simple energetic argument within
the simplified description of anisotropic magnetohydrodynamics \cite{PRS06}. 
}

The linear mirror instability has been extensively studied both analytically
(see e.g. \cite{PokhotelovSagdeev,Hell07}), and by means of
particle-in-cell (PIC) simulations \cite{G92}. As shown in \cite%
{hasegawa,hall,PokhotelovSagdeev}, the instability is arrested at large $k$
due to finite ion Larmor radius (FLR) effects. It turns out that
wave-particle resonance plays a central role in driving the instability,
while the FLR effects are at the origin of the quenching of the instability
at small scales. In contrast, a few years ago,  a theoretical understanding of
the nonlinear phase remained limited to phenomenological modeling of
particle trapping \cite{KS96,P98} that hardly reproduce simulations of
Vlasov-Maxwell equations \cite{BSD03}.

The first nonlinear theory was formulated in  \cite{KPS2007a,
KPS2007b} where we developed a weakly
nonlinear approach to the mirror instability based on the mixed
hydrodynamic-kinetic description. For the sake of simplicity, 
an electron-proton plasma with cold electrons was considered first.
It includes the force-balance equation
within the anisotropic MHD and the drift kinetic equation for the  ions. Close to
threshold, the unstable modes have wavevectors almost perpendicular to the
ambient magnetic field $\mathbf{B}$ ($k_{z}/k_{\perp }\ll 1$) with $k_{\perp
}\rho _{i}\ll 1$ ($\rho _{i}$ is the ion Larmor radius), so that the
perturbations can be described using a long-wave approximation. The latter
allows one to apply the drift kinetic equation (see, e.g. \cite{sivukhin,
kulsrud}) to estimate the main nonlinear effects that correspond to a local
shift of the instability threshold (\ref{thr}). All other nonlinearities
connected, for example, with ion inertia are smaller. As the result, it is
possible to derive an asymptotic equation with quadratic nonlinearity of 
generalized gradient type \cite{KPS2007a,KPS2007b}. The latter property
implies an irreversible character of the mirror modes behavior, associated
with ion Landau damping, where the free energy (or Lyapunov functional) can
only decrease in time. In this framework, above threshold, the mirror
modes have a blow-up behavior with a possible saturation at an amplitude
level comparable to that of the ambient field. Below threshold, all
stationary (localized) structures were predicted to be unstable. Thus, the
system near the  MI threshold displays a  behavior typical of a
subcritical bifurcation when the small-amplitude stationary solutions below
threshold turn out to be unstable; above threshold, the amplitude of magnetic
field perturbations tends to blow up. It is worth noting that this approach
contrasts with the quasi-linear theory \cite{ShSh64} that also assumes
vicinity of the instability threshold but, being based on a random phase
approximation, cannot predict the appearance of coherent structures.
Phenomenological models based on the cooling of trapped particles were
proposed to interpret the existence of deep magnetic holes \cite{KS96,Pant98}
These models do not however address the initial value problem in the
mirror unstable regime.

The asymptotic model \cite{KPS2007a, KPS2007b} was first derived under the
assumption of  cold electrons.  Therefore,  in our further papers \cite%
{KPS2012a, KPS2012b},  we considered how hot electrons can be incorporated
into the model. The approach we developed is based on the assumption of an 
adiabatically slow dynamics of the mirror structures that allows one to
compute the coupling coefficient in the weakly nonlinear regime
as well as to simplify all calculations of the linear growth rate in the
case of bi-Maxwellian distributions for both the ions and the electrons. The
adiabatic hypothesis can be proved perturbatively, and is in particular
valid within the asymptotic model. Because this model predicts the
existence of subcritcal bifurcation with a blow-up behavior above threshold, 
consistent with the formation of mirror structures with amplitude of the magnetic
field perturbation comparable with the ambient field, our next step was to
investigate the properties of possible stationary mirror structures.

The aim of the present paper is twofolds. The first part provides a 
review of  our previous results concerning the weakly nonlinear model for
both cold (\cite{KPS2007a, KPS2007b}) and hot (\cite{KPS2012a}) electrons.
Another goal of this paper is to study steady mirror structures resulting
from the balance of magnetic and (both parallel and perpendicular) thermal
pressures, whose simplest description is provided by anisotropic MHD.
Isotropic MHD equilibria are classically governed by the Grad-Shafranov (GS)
equation \cite{grad, shafranov-58, shafranov-66}. We here revisit this
approach in the case of anisotropic electron and ion fluids where the
perpendicular and parallel pressures are given by equations of state
appropriate for the static character of the solutions. However, the MHD
stationary equations, at least in the two-dimensional geometry, turn out to
be ill-posed. As a consequence, these equations require some regularization.
As done in a similar context of pattern formation \cite%
{ErcolaniIndikNewellPassot}, an additional linear term involving a square
Laplacian is added. For nonlinear mirror modes, regularization can originate
from finite Larmor radius (FLR) corrections, which are not retained in the
present analysis based on the drift kinetic equation (see, e.g. \cite%
{KPS2007a, KPS2007b}).

The paper is organized as follows. In Section II, we discuss the linear
mirror instability near the
MI threshold. Section III is devoted to the derivation of weakly nonlinear
asymptotic model, in the simplest case of cold electrons, and to
its properties, including possible stationary states
(below the MI threshold) and blow-up behavior (above threshold). Section IV
deals with accounting electrons in the asymptotic model. Here we develop the
adiabatic approach for finding contributions from electrons to both the linear
growth rate and the nonlinear coupling coefficient entering the asymptotic model. In Section V,
we  formulate the variational principle for the stationary anisotropic 
MHD when both parallel and transverse pressures depend on the magnetic 
field amplitude with a free energy given by the space integral of the
parallel tension. In this case, as well known \cite{Taylor1963, Taylor1964,
NorthropWhiteman1964, Grad1967, HallMcNamara1975, ZakharovShafranov}, the
parallel component of the MHD equation is satisfied identically.  
In Section 6, the anisotropic Grad-Shafranov equations are revisited when
the gyrotropic pressures depend only on the local magnetic field amplitude
that, as shown in the forthcoming sections, is specific of  nonlinear mirror
modes. In this case  the stationary anisotropic MHD represents 
an  hydrodynamic integrable-type system and for
this reason requires the renormalization due to  FLR effects.
In this Section, it is shown also that the equations of state
resulting from an adiabatic approximation of the drift kinetic description,
require a regularization because of an overestimate of the contributions from
the particles with a large magnetic moment. We discuss in particular the
small-amplitude regime and show that the pressure-balanced structures are
then governed by the KPII equation which possesses lump solutions. Numerical
simulations reproduce these special structures, that turn out to be
unstable. Computation of stable solutions lead to large-amplitude purely
one-dimensional solutions in the form of stripes that appear to be sensitive
to the regularization process, an indication that the regime cannot be
captured by the drift kinetic approximation and that finite Larmor
corrections and trapped particles are to be retained.  Section VII aims for presentation of the
numerical results for two-dimensional (depending on $x$ and $y$ coordinates)
stationary mirror structures when the magnetic field $\mathbf{B}$ has also a $%
B_{z}$ component. In particular, we show that for small $B_{z}$ stationary
structures realizing the minimum of the free energy, below and above the
threshold, have the form of stripes which are one-dimensional structures
with constant magnetic field outside and inside the stripes. The transient
region, between outer and inner regions, for the stripes represents the
magnetic well which structure is defined by the FLR contributions to the
free energy. With increasing $B_{z}$, instead of stripes, the free energy has
its minimum for  bubble-type structures with an  elliptic form. When $%
B_{x,y}\rightarrow 0$ these bubbles become circular. In this case,  FLR
effects play a role of the surface tension. Section VIII is the
conclusion.

\section{Main equations and mirror instability}

Consider for the sake of simplicity, a plasma with cold electrons. To
describe the mirror instability in the long-wave limit it is enough to use
the drift kinetic equation for ions ignoring parallel electric field $%
E_{\Vert }$ and transverse electric drift: 
\begin{equation}
\frac{\partial f}{\partial t}+v_{\Vert }\mathbf{b}\cdot \nabla f-\mu \mathbf{%
b}\cdot \nabla B\frac{\partial f}{\partial v_{\Vert }}=0\mathbf{.}
\label{mainkin}
\end{equation}
In this approximation ions move along the magnetic field ($\mathbf{b}=%
\mathbf{B}/B$) due to the magnetic force $\mu \mathbf{\ b}\cdot \nabla B$
where $\mu =v_{\perp }^2/(2B)$ is the adiabatic invariant which plays the
role of a parameter in equation (\ref{mainkin}). Both pressures $p_{\Vert }$
and $p_{\perp }$ are given by 
\begin{equation}
p_{\Vert }=mB\int v_{\Vert }^2fd\mu dv_{\Vert }d\varphi \equiv m\int
v_{\Vert }^2fd^3v,  \label{par}
\end{equation}
\begin{equation}
p_{\perp }=mB^2\int \mu fd\mu dv_{\Vert }d\varphi \equiv \frac 12m\int
v_{\perp }^2fd^3v.  \label{perp}
\end{equation}
Equation (\ref{mainkin}) with relations (\ref{par}) and (\ref{perp}) are
supplemented with the equation expressing the balance of forces in a plane
transverse to the local magnetic field 
\begin{eqnarray}
&&{\hat\Pi} \Big \{ -\nabla \Big( p_{\perp }+ \frac{B^2}{8\pi }\Big) 
\nonumber \\
&&+ \Big[ 1+\frac{4\pi }{B^2}\Big( p_{\perp}-p_{\Vert }\Big) \Big] \frac{%
\mathbf{B}\cdot \nabla \mathbf{B}}{4\pi } \Big \}=0.  \label{balance-nl}
\end{eqnarray}
Here, consistently with the long-wave approximation, we neglect both the
plasma inertia and the non-gyrotropic contributions to the pressure tensor.
Furthermore, ${\hat\Pi} _{ik}=\delta_{ik}-b_ib_k$ denotes the projection
operator in the plane transverse to the local magnetic field. In this
equation, the first term describes the action of the magnetic and
perpendicular pressures, the second term being responsible for magnetic
lines elasticity.

The equation governing the mirror dynamics is then obtained perturbatively
by expanding Eqs. (\ref{mainkin}) and (\ref{balance-nl}). In this approach,
the ion pressure tensor elements are computed from the system (\ref{mainkin}%
), (\ref{balance-nl}), near a bi-Maxwellian equilibrium state characterized
by temperatures $T_{\perp }$ and $T_{\Vert }$ and a constant ambient
magnetic field $\mathbf{B_0}$ taken along the $z$-direction.

From Eq. (\ref{balance-nl}) linearized about the background field $\mathbf{%
B_0}$ by writing $\mathbf{B=B}_0+\widetilde{\mathbf{B}}$ ($B_0\gg \widetilde{%
B}$) with $\widetilde{\mathbf{B}}\sim e^{-i\omega t+i\mathbf{k\cdot r}}$, we
have 
\begin{equation}
p_{\perp }^{(1)}+\frac{B_0\widetilde{B}_z}{4\pi }=-\frac{k_{z}^2}{k_{\perp
}^2}\left( 1+\frac{\beta _{\perp }-\beta _{\Vert }}2\right) \frac{B_0%
\widetilde{B}_z}{4\pi }.  \label{first}
\end{equation}
Here $k_z$ and $k_{\perp }$ are the projections of the wave vector $\mathbf{k%
}$, and $p_{\perp }^{(1)}$ is calculated from the linearized drift kinetic
equation (\ref{mainkin}): 
\[
\frac{\partial f^{(1)}}{\partial t}+v_{\Vert }\frac{\partial f^{(1)}}{%
\partial z}-\mu \frac{\partial \widetilde{B}_z}{\partial z}\frac{\partial
f^{(0)}}{\partial v_{\Vert }}=0. 
\]
In Fourier space, this equation has the solution 
\begin{equation}
f^{(1)}=-\frac{\mu \widetilde{B}_z}{\omega -k_zv_{\Vert }}k_z\frac{\partial
f^{(0)}}{\partial v_{\Vert }}.  \label{omega-kin}
\end{equation}
The mirror instability is such that $\omega /k_z\ll v_{th{\Vert }}=\sqrt{%
2T_{\Vert }/m}$. This means that the ions contributing to the resonance $%
\omega -k_zv_{\Vert}=0$, correspond to the maximum of the ion distribution
function.

After substituting (\ref{omega-kin}) into the first order term for
perpendicular pressure (\ref{perp}) and performing integration, we get 
\begin{equation}
p_{\perp }^{(1)}=\beta _{\perp }\left( 1-\frac{\beta _{\perp }}{\beta
_{\Vert }}\right) \frac{B_0\widetilde{B}_z}{4\pi }-\frac {i\sqrt{\pi }\omega%
} {|k_z|v_{th{\Vert }}}\frac{\beta _{\perp }^2}{\beta _{\Vert }}\frac{B_0%
\widetilde{B}_z}{4\pi }.  \label{first-perp}
\end{equation}
The first term in (\ref{first-perp}) is due to the difference between
perpendicular and parallel pressures, while the second one accounts for the
Landau pole.

Equation (\ref{first-perp}) together with (\ref{first}) yield the growth
rate for the mirror instability in the drift approximation where FLR
corrections are neglected \cite{VedenovSagdeev} 
\begin{equation}
\gamma =|k_{z}|v_{th{\Vert }}\frac{\beta _{\Vert }}{\sqrt{\pi }\beta _{\perp
}}\left[ \frac{\beta _{\perp }}{\beta _{\Vert }}-1-\frac{1}{\beta _{\perp }}-%
\frac{k_{z}^{2}}{k_{\perp }^{2}\beta _{\perp }}\chi \right] ,  \label{growth}
\end{equation}%
where $\chi =1+(\beta _{\perp }-\beta _{\Vert })/2$. The instability takes
place when the criterion (\ref{thr}) is fulfilled and, near threshold,
develops in quasi-perpendicular directions, making the parallel magnetic
perturbation dominant.

As shown in Refs. \cite{hasegawa, hall, PokhotelovSagdeev}, when the FLR
corrections are relevant, the growth rate is modified into 
\begin{equation}
\gamma =|k_{z}|v_{th{\Vert }}\frac{\beta _{\Vert }\chi }{\sqrt{\pi }\beta
_{\perp }^{2}}\left[ \varepsilon -\frac{k_{z}^{2}}{k_{\perp }^{2}}-\frac{3}{%
4\chi }k_{\perp }^{2}\rho _{i}^{2}\right]  \label{gamma}
\end{equation}%
where $\varepsilon =\beta _{\perp }\chi ^{-1}(\beta _{\perp }/\beta _{\Vert
}-1-\beta _{\perp }^{-1})$ and the ion Larmor radius $\rho _{i}=v_{th\perp
}/\omega _{ci}$ is defined with the transverse thermal velocity $v_{th\perp
}=\sqrt{2T_{\perp }/m}$ and the ion gyrofrequency $\omega _{ci}=eB_{0}/(mc)$%
. This growth rate can be recovered by expanding the general expression
given in \cite{PokhotelovSagdeev}, in the limit of small transverse
wavenumbers. It can also be obtained directly from the Vlasov-Maxwell (VM)
equations in a long-wave limit which retains non gyrotropic contributions 
\cite{Calif08}. It is important to note that the expression (\ref{gamma})
for $\gamma $ is consistent with the applicability condition $\omega
/k_{z}\ll v_{th{\Vert }}$, i.e. when the supercritical parameter $%
|\varepsilon |\ll 1$. In this case the instability saturation happens at
small $k_{\perp }\propto \sqrt{\varepsilon }$ due to FLR and for almost
perpendicular direction in a small cone of angles, $k_{z}/$ $k_{\perp
}\propto \sqrt{\varepsilon }$. As a result, the growth rate $\gamma \propto
\varepsilon ^{2}$, so that, when defining new stretched variables by 
\begin{eqnarray}
&&k_{z}=\varepsilon K_{z}\rho _{i}^{-1}(2/\sqrt{3})\chi ^{1/2},  \nonumber \\
&&k_{\perp }=\sqrt{\varepsilon }(2/\sqrt{3})K_{\perp }\rho _{i}^{-1}\chi
^{1/2},  \label{scaling} \\
&&\gamma =\varepsilon ^{2}\Gamma (2/\sqrt{3})\Omega \left( \sqrt{\pi }\beta
_{\perp }\right) ^{-1}\left( {\chi }\beta _{\Vert }/\beta _{\perp }\right) {%
^{3/2}},  \nonumber
\end{eqnarray}%
it takes the form 
\begin{equation}
\Gamma =|K_{z}|\left( 1-{K_{z}^{2}}/{K_{\perp }^{2}}-K_{\perp }^{2}\right) .
\label{Gamma}
\end{equation}%
Hence it is seen that, in the ($K_{\perp }-\Theta $) plane ($\Theta \equiv
K_{z}/K_{\perp }$), the instability takes place inside the unit circle: $%
\Theta ^{2}+K_{\perp }^{2}<1.$ The maximum of $\Gamma $ is obtained for $%
K_{\perp }=1/2$, $\Theta =\pm $ $1/2$ and is equal to $\Gamma _{\max }=1/8$.
Outside the circle, the growth rate becomes negative (in agreement with \cite%
{hall}).

\section{\protect\bigskip Weakly nonlinear regime: asymptotic model for cold
electrons}

\subsection{Derivation}

As it follows from (\ref{first}), in the linear regime, near the instability
threshold, the fluctuations of perpendicular and magnetic pressures almost
compensate each other (compare with (\ref{growth})). Therefore, in the
nonlinear stage of this instability, we can expect that the main nonlinear
contributions come from the second order corrections to the total
(perpendicular plus magnetic) pressure, i.e. 
\begin{equation}
p_{\perp }^{(1)}+\frac{B_{0}\widetilde{B}_{z}}{4\pi }+p_{\perp }^{(2)}+\frac{%
\widetilde{B}_{z}^{2}}{8\pi }=-\chi \frac{\partial _{z}^{2}}{\Delta _{\perp }%
}\frac{B_{0}\widetilde{B}_{z}}{4\pi }.  \label{secondorder}
\end{equation}%
This result can be obtained rigorously by means of a multi-scale expansion
based on the linear theory scalings (\ref{scaling}). For this purpose, we
introduce a slow time $T$ and slow coordinates $\mathbf{R}$ in a way
consistent with (\ref{scaling}), and expand the magnetic field fluctuations
as a powers series in $\varepsilon ^{1/2}$: 
\begin{equation}
\widetilde{B}_{z}=\varepsilon {B}_{z}^{(1)}+O(\varepsilon ^{2}),\,\,\,%
\widetilde{\mathbf{B}}_{\perp }=\varepsilon ^{3/2}\mathbf{B}_{\perp
}^{(3/2)}+O(\varepsilon ^{5/2}),  \label{slow-magnetic}
\end{equation}%
where $\mathbf{B}^{(n/2)}$ are assumed to be functions of $\mathbf{R}$ and $%
T $. Using these expressions, it is easy to establish that quadratic
nonlinear terms coming from the expansion of $\Pi $ in (\ref{balance-nl}) as
well as from the second term in the r.h.s. of Eq. (\ref{secondorder}) are
small in comparison with the quadratic term originating from the magnetic
pressure in Eq. (\ref{secondorder}). 
Thus, to get a nonlinear model for mirror dynamics, it is enough to find $%
p_{\perp }^{(2)}$. The expansion (\ref{slow-magnetic}) induces a
corresponding expansion for the distribution function and for both
pressures. Defining 
\[
{\widetilde{p}}_{\perp }^{(n)}=\pi m\int v_{\perp }^{2}f^{(n)}v_{\perp
}dv_{\perp }dv_{\Vert }, 
\]%
from (\ref{perp}) we have 
\[
p_{\perp }^{(2)}=(B_{z}^{(1)}/{B_{0}})^{2}p_{\perp }^{(0)}+2({B_{z}^{(1)}}/{%
B_{0}})\,\widetilde{p}_{\perp }^{(1)}+\widetilde{p}_{\perp }^{(2)}, 
\]%
up to an additional contribution proportional to $B_{z}^{(2)}$ that cancels
out in the final equation due to the threshold condition.

On the considered time scale, the effect of nonlinear Landau resonance is
negligible in the contribution to $f^{(2)}$ that can thus be estimated from
the equation 
\[
v_{\Vert }\frac{\partial f^{(2)}}{\partial z}+({2\mu ^{2}}/{v_{th\Vert }^{2}}%
)B_{z}^{(1)}\frac{\partial B_{z}^{(1)}}{\partial z}\frac{\partial f^{(0)}}{%
\partial {v_{\Vert }}}=0. 
\]%
For an equilibrium bi-Maxwellian distribution, we have 
\begin{equation}  \label{second-order}
f^{(2)}=(2\mu ^{2}/v_{th\Vert }^{4})(B_{z}^{(1)})^{2}f^{(0)}
\end{equation}
and thus 
\[
p_{\perp }^{(2)}=\left( \beta _{\perp }-4{\beta _{\perp }^{2}}/{\beta
_{\Vert }}+3{\beta _{\perp }^{3}}/{\beta _{\Vert }^{2}}\right) \frac{%
\widetilde{B}_{z}^{2}}{8\pi }. 
\]%
As a consequence, because of the vicinity to threshold we obtain 
\begin{equation}
p_{\perp }^{(2)}+\frac{\widetilde{B}_{z}^{2}}{8\pi }=\left( 1+{\beta _{\perp
}^{-1}}\right) \frac{3{\widetilde{B}_{z}^{2}}}{8\pi }>0.
\label{p-perp-second}
\end{equation}%
Then rewriting equation (\ref{secondorder}) using the slow variables (\ref%
{scaling}) and rescaling the amplitude 
\[
{\widetilde{B}_{z}}/{B_{0}}=\varepsilon {2\chi \beta _{\perp }}(1+\beta
_{\perp })^{-1}u, 
\]%
we arrive at the equation \cite{KPS2007a, KPS2007b} 
\begin{equation}
\frac{\partial u}{\partial T}=\widehat{K}_{Z}\left[ \left( \sigma -\Delta
_{\perp }^{-1}\frac{\partial ^{2}}{\partial Z^{2}}+\Delta _{\perp }\right)
u-3u^{2}\right] .  \label{main}
\end{equation}%
Here $\sigma =\pm 1$, depending of the positive or negative sign of $%
\varepsilon $, $\widehat{K}_{Z}=-\mathcal{H}\partial _{Z}$ is a positive
definite operator (whose Fourier transform is $|K_{Z}|$), $\widehat{H}$ is
Hilbert transform: 
\[
\widehat{H}f(Z)=\frac{1}{\pi }VP\int_{-\infty }^{\infty }\frac{f(Z^{\prime })%
}{Z^{\prime }-Z}dZ^{\prime }. 
\]%
As seen from the equation, its linear part reproduces the growth rate (\ref%
{Gamma}). In particular, the third term in the r.h.s. accounts for the FLR
effect.

Equation (\ref{main}) simplifies when the spatial variations are limited to
a direction making a fixed angle with the ambient magnetic field. After a
simple rescaling, one gets 
\begin{equation}
\frac{\partial u}{\partial T}=\widehat{K}_{\Xi }\left[ \left( \sigma +\frac{%
\partial ^{2}}{\partial \Xi ^{2}}\right) u-3u^{2}\right] ,  \label{oneD}
\end{equation}%
where $\Xi $ is the coordinate along the direction of variation. This
equation can be referred to as a \textquotedblleft dissipative Korteveg-de
Vries (KdV) equation\textquotedblright , since its stationary solutions
coincide with those of the usual KdV equation. The presence of the Hilbert
transform in Eq. (\ref{oneD}) nevertheless leads to a dynamics significantly
different from that described by soliton equations. Besides, it is worth
noting also that Eq. (\ref{main}) in the two-dimensional case has some
similarity with the KP equation (see, Section IV).

\subsection{Properties of the asymptotic model}

Equation (\ref{main}) (and its 1D reduction (\ref{oneD}) as well) possesses
the remarkable property of being of the form 
\[
\frac{\partial u}{\partial T}=-\widehat{K}_{z}\frac{\delta F}{\delta u}, 
\]%
where 
\begin{eqnarray}
F &=&\int \left[ -\frac{\sigma }{2}u^{2}+\frac{1}{2}u\Delta _{\perp
}^{-1}\partial _{Z}^{2}u+\frac{1}{2}\left( \nabla _{\perp }u\right)
^{2}+u^{3}\right] d\mathbf{R}  \nonumber \\
\ &\equiv &-\sigma N/2+I_{1}/2+I_{2}/2+I_{3}  \label{free}
\end{eqnarray}%
has the meaning of a free energy or a Lyapunov functional. This quantity can
only decrease in time, since 
\begin{equation}
\frac{dF}{dt}=\int \frac{\delta F}{\delta u}\frac{\partial u}{\partial t}d%
\mathbf{R}=-\int \frac{\delta F}{\delta u}\widehat{K}_{z}\frac{\delta F}{%
\delta u}d\mathbf{R}\leq 0.  \label{F-time}
\end{equation}%
This derivative can only vanish at the stationary localized solutions,
defined by the equation 
\begin{equation}
\frac{\delta F}{\delta u}=\left( \sigma -\Delta _{\perp }^{-1}\frac{\partial
^{2}}{\partial Z^{2}}+\Delta _{\perp }\right) u-3u^{2}=0.  \label{stat}
\end{equation}

We now show that non-zero solutions of this equation do not exist above
threshold ($\sigma =+1$). For this aim, following Ref. \cite{kuzn-musher},
we establish relations between the integrals $N$, $I_{1}$, $I_{2}$ and $%
I_{3} $, using the fact that solutions of Eq. (\ref{stat}) are stationary
points of the functional $F$ (i.e. $\delta F=0$). Multiplying Eq. (\ref{stat}%
) by $U $ and integrating over $\mathbf{R}$ gives the first relation 
\[
\sigma N-I_{1}-I_{2}-3I_{3}=0. 
\]%
Two other relations can be found if one makes the scaling transformations, $%
Z\rightarrow aZ,$ $\mathbf{R}_{\perp }\rightarrow b\mathbf{R}_{\perp }$,
under which the free energy (\ref{free}) becomes a function of two scaling
parameters $a$ and $b$ 
\[
F(a,b)=-\frac{\sigma N}{2}ab^{2}+\frac{I_{1}}{2}b^{4}a^{-1}+\frac{I_{2}}{2}%
a+I_{3}ab^{2}. 
\]%
Due to the condition $\delta F=0$, the first derivatives of $F$ at $a=b=1$
have to vanish: 
\begin{eqnarray}
\frac{\partial F}{\partial a}=-\frac{\sigma N}{2}-\frac{I_{1}}{2}+\frac{I_{2}%
}{2}+I_{3} &=&0,  \nonumber \\
\frac{\partial F}{\partial b}=-\sigma N+2I_{1}+2I_{3} &=&0.  \nonumber
\end{eqnarray}%
Hence, after simple algebra, one gets the three relations 
\[
I_{1}+\frac{\sigma }{2}N=0,\,\,\,I_{3}=-2I_{1},\,\,\,I_{2}=3I_{1}. 
\]%
For $\sigma =+1$, the first relation can be satisfied only by the trivial
solution $u=0$, because both integrals $I_{1}$ and $N$ are positive
definite. In other words, above threshold, nontrivial stationary solutions
obeying the prescribed scalings do not exist.

In contrast, below threshold, stationary localized solutions can exist. For
these solutions, the free energy is positive and reduces to $F_s=N/2$.
Furthermore, $I_3=\int U^3 d^3R<0$ which means that the structures have the
form of magnetic holes. As stationary points of the functional $F$, these
solutions represent saddle points, since the corresponding determinant of
second derivatives of $F$ with respect to scaling parameters taken at these
solutions is negative ($\partial_{aa}
F\partial_{bb}F-(\partial_{ab}F)^2=-2N^2<0$). As a consequence, there exist
directions in the eigenfunction space for which the free-energy
perturbation is strictly negative, corresponding to linear instability of
the associated stationary structure. This is one of the properties for
subcritical bifurcations.

As a consequence, starting from general initial conditions, the derivative $%
dF/dt$ (\ref{F-time}) is almost always \textit{negative}, except for
unstable stationary points (zero measure) below threshold. In the nonlinear
regime, negativeness of this derivative implies $\int u^{3}d^{3}R<0$, which
corresponds to the formation of magnetic holes. Moreover, this nonlinear
term (in $F$) is responsible for collapse, i.e. formation of singularity in
a finite time.

\subsection{\protect\bigskip Blow-up}

In order to characterize the nature of the singularity of Eq. (\ref{oneD}),
it is convenient to introduce the similarity variables $\xi
=(T_{0}-T)^{-1/3}\Xi $, $\ \tau =-\log (T_{0}-T)$, and to look for a
solution in the form $U=\left( T_{0}-T\right) ^{-2/3}g(\xi ,\tau )$, where $%
g(\xi ,\tau )$ satisfies the equation 
\[
\frac{\partial g}{\partial \tau }+\frac{2}{3}g+\frac{\xi }{3}\frac{\partial g%
}{\partial \xi }=\widehat{K}_{\xi }\left[ \frac{\partial ^{2}g}{\partial \xi
^{2}}-3g^{2}\right] +e^{-\tau }\widehat{K}_{\xi }g. 
\]%
As time $T$ approaches $T_{0}$ ($\tau \rightarrow \infty $), the last term
in this equation becomes negligibly small and simultaneously $\partial
_{\tau }g\rightarrow 0$ so that asymptotically the equation transforms into 
\begin{equation}
\frac{2}{3}g+\frac{\xi }{3}\frac{dg}{d\xi }=\widehat{K}_{\xi }\left[ \frac{%
d^{2}g}{d\xi ^{2}}-3g^{2}\right] .  \label{asympEq}
\end{equation}%
For the free energy this means that close to $T_{0}$ the first term $\sim N$
turns out to be much smaller in comparison with all other contributions, in
particular with $\int U^{3}d\Xi $.

At large $|\xi |$, that corresponds to the limit $T\to T_0$, the asymptotic
solution $\widetilde{g}$ of Eq. (\ref{asympEq}) obeys 
\[
2\widetilde{g}+\xi \frac{d \widetilde{g}}{d\xi} =C\xi ^{-2} 
\]
where $C=\frac 9\pi \int_{-\infty }^\infty g^2(\xi ^{\prime })d\xi ^{\prime
}>0$, and has the form $\widetilde{g}{=C\xi ^{-2}\log \left| \xi /\xi
_0\right| }$. For $U$, it gives the asymptotic solution 
\[
U_{asymp}=\frac{C}{\Xi^2}\log | \Xi / \Xi_0(t)| 
\]
with $\displaystyle{\Xi_0(t)=(T_0-T)^{1/3} \xi_0}$, that, as $T\to T_0$, has
an almost time independent tail. For $|\Xi |<\left( T_0-T\right)
^{1/3}|\xi_0|$, the solution is negative and becomes singular as $\Xi $
approaches the origin.

Asymptotically self-similar solutions can also be constructed in three
dimensions, when rescaling the longitudinal coordinate by $(T_{0}-T)^{1/2}$,
the transverse ones by $(T_{0}-T)^{1/4}$ and the amplitude of the solution
by $(T_{0}-T)^{-1/2}$. Existence of a finite time singularity for the
initial value problem can be established for initial conditions for which
the functional $F$ is negative, when the term involving $\sigma $ can be
neglected, an approximation consistent with the dynamics: 
\begin{equation}
F\rightarrow F_{\lim }\equiv \frac{I_{1}}{2}+\frac{I_{2}}{2}+I_{3}.
\label{F-lim}
\end{equation}%
To prove this statement, consider the operator $\widehat{K}_{z}^{-1}$,
(inverse of the operator $\widehat{K}_{z}$), which is defined on functions
obeying $\int U(Z,\mathbf{R}_{\perp })dZ=0$, a condition consistent with Eq.
(\ref{main}). Then the time derivative of $F_{\lim }$ can be rewritten
through the operator $\widehat{K}_{z}^{-1}$ as follows, 
\begin{equation}
\frac{dF_{\lim }}{dT}=-\int U_{T}\widehat{K}_{z}^{-1}U_{T}d\mathbf{R}\leq 0.
\label{F-reduce}
\end{equation}%
Consider now the positive definite quantity $\widetilde{N}=\int U\widehat{K}%
_{z}^{-1}Ud\mathbf{R}\geq 0,$ whose dynamics is determined by the equation 
\begin{equation}
\frac{d\widetilde{N}}{dT}=-2\left( I_{1}+I_{2}+3I_{3}\right) =-6F_{\lim
}+I_{1}+I_{2}.  \label{N-t}
\end{equation}%
Let $F_{\lim }$ be negative initially, then at $T\geq 0$ the r.h.s. of (\ref%
{N-t}) will be positive, and, as a consequence, $\widetilde{N}$ will be a
growing function of time.

Introduce now the new quantity $S=-F_{lim}/ \widetilde{N}$ which is positive
definite if $F_{lim}|_{T=0}<0$. The time derivative of $S$ is then defined
by means of Eqs. (\ref{F-reduce}) and (\ref{N-t}): 
\begin{equation}
\frac{dS}{dT}=-\frac{F_{\lim }\widetilde{N}_T}{\widetilde{N}^2}+\frac 1{%
\widetilde{N}}\int U_T\widehat{K}_z^{-1}U_Td\mathbf{R}.  \label{S_t}
\end{equation}
The second term in the r.h.s. of this equation can be estimated using the
Cauchy-Bunyakowsky inequality: 
\[
\frac{d\widetilde{N}}{dT}=2\int U\widehat{K}_z^{-1}U_Td\mathbf{R}\leq 2%
\widetilde{N}^{1/2}\left( \int U_T\widehat{K}_z^{-1}U_Td\mathbf{R}\right)
^{1/2}, 
\]
that gives 
\[
\int U_T\widehat{K}_z^{-1}U_Td\mathbf{R}\geq {\widetilde{N}_T^2}/({4%
\widetilde{N}}). 
\]
Substituting the obtained estimate into Eq. (\ref{S_t}) and taking into
account definition (\ref{F-lim}) for $F_{\lim }$ and Eq. (\ref{N-t}) as
well, we arrive at the differential inequality for $S$ (compare with \cite%
{Turitsyn}): 
\[
\frac{dS}{dT}\geq \frac{\widetilde{N}_T}{\widetilde{N}^2}\left[ \frac{%
\widetilde{N}_T}4-F_{\lim }\right] \geq 15\,S^2. 
\]
Integrating this first-order differential inequality yields 
\begin{equation}
S\geq \frac 1{15(T_0-T)}.  \label{crit}
\end{equation}
Here the collapse time $T_0=(15\,S_0)^{-1}$ is expressed in terms of the
initial value $S|_{t=0}=S_0$. It is interesting to mention that the time
behavior of $S$ given by the estimate (\ref{crit}) coincides with that given
by the self-similar asymptotics.

\subsection{Conclusion of Section III}

We have presented an asymptotic description of the nonlinear dynamics
of mirror modes near the instability threshold. Below threshold, we have
demonstrated the existence of unstable stationary solutions. Differently,
above threshold, no stationary solution consistent with the prescribed
small-amplitude, long-wavelength scaling can exist. For small-amplitude
initial conditions, the time evolution predicted by the asymptotic equation (%
\ref{main}) leads to a finite-time singularity. These properties are based
on the fact that this equation belongs to the generalized gradient systems
for which it is possible to introduce a free energy or a Lyapunov functional
that decreases in time.

The singularity formation as well as the existence of unstable stationary
structures below the mirror instability threshold obtained with the
asymptotic model, can be viewed as features of a subcritical bifurcation
towards a large-amplitude state that cannot be described in the framework of
the present analysis. Such an evolution should indeed involve saturation
mechanisms that become relevant when the perturbation amplitudes become
comparable with the ambient field.

\section{Adiabatic approach: Account of electrons}

The mirror instability, as known, is a kinetic instability whose growth rate
was first obtained under the assumption of cold electrons \cite%
{VedenovSagdeev}, a regime where the contributions of the parallel electric
field $E_{\Vert }$ can be neglected. However, in realistic space plasmas,
the electron temperature can hardly be ignored \cite{Stverak08}. The linear
theory retaining the electron temperature and its possible anisotropy, in
the quasi-hydrodynamic limit (which neglects finite Larmor radius
corrections), was developed in the case of bi-Maxwellian distribution
functions by several authors (see e.g. \cite{Stix62}, \cite%
{PokhotelovSagdeev}, \cite{Hell07}). A general estimate of the growth rate
under the sole condition that it is small compared with the ion
gyrofrequency (a condition reflecting close vicinity to threshold) is
presented in \cite{KPS2012a}. Like for the cold electrons case, the
instability develops in quasi-perpendicular directions, making the parallel
magnetic perturbation dominant. This analysis includes in particular regimes
with a significant electron temperature anisotropy for which the instability
extends beyond the ion Larmor radius. In the limit where the instability is
limited to scales large compared with the ion Larmor radius $\rho_i$, only
the leading order contribution in terms of the small parameter $\gamma
/(|k|_{z}v_{\Vert i})$ is to be retained in estimating Landau damping, and
the growth rate is given by 
\begin{eqnarray}
&&\gamma =\frac{2}{\sqrt{\pi }}\frac{T_{\Vert i}}{T_{\perp i}}\frac{%
|k_{z}|v_{\Vert i}}{E}\Big \{\Gamma -\frac{1}{\beta _{\perp }}\Big (1+\frac{%
\beta _{\perp }-\beta _{\Vert }}{2}\Big )\frac{k_{z}^{2}}{k_{\perp }^{2}} 
\nonumber \\
&&\qquad -\frac{3}{4(1+\theta _{\perp })}\Big (\frac{T_{\perp i}}{T_{\Vert i}%
}-1\Big )(1+F)k_{\perp }^{2}r_{L}^{2}\Big \},  \label{growth_rate}
\end{eqnarray}%
where 
\begin{equation}
\Gamma =\frac{T_{\perp i}}{T_{\parallel i}}\frac{(\theta _{\parallel
}+\theta _{\perp })^{2}+2\theta _{\parallel }(\theta _{\perp }^{2}+1)}{%
2\theta _{\parallel }(1+\theta _{\perp })(\theta _{\parallel }+1)}-1-\frac{1%
}{\beta _{\perp }}  \label{newthreshold}
\end{equation}%
measures the distance to threshold and 
\begin{eqnarray}
E &=&\frac{1+\theta _{\perp }}{(1+\theta _{\parallel })^{2}}\left[ 2+\theta
_{\perp }(4+\theta _{\perp })+\theta _{\parallel }^{2}\right]  \nonumber \\
F &=&\frac{T_{\Vert e}}{T_{\Vert e}+T_{\Vert i}}\Big \{-1+\frac{\theta
_{\perp }}{\theta _{\Vert }}  \nonumber \\
&&-\frac{2}{3}\frac{T_{\Vert i}}{T_{\perp i}}\Big [\Big (\frac{T_{\Vert i}}{%
T_{\perp i}}-1\Big )\frac{1}{\beta _{\perp i}}-\theta _{\perp }\Big (\frac{%
T_{\perp e}}{T_{\Vert e}}-1\Big )\Big ]\Big \}.  \nonumber
\end{eqnarray}%
Here, ${T_{\perp \alpha }}$ and ${T_{\Vert \alpha }}$ are the perpendicular
and parallel (relative to the ambient magnetic field $\mathbf{B}_{0}$ taken
in the $z$ direction) temperatures of the species $\alpha $ ($\alpha =i$ for
ions and $\alpha =e$ for electrons ), $\theta _{\perp }={T_{\perp e}}/{%
T_{\perp i}}$, $\theta _{\Vert }={T_{\Vert e}}/{T_{\Vert i}}$ and $\beta
_{\perp }=\beta _{\perp i}+\beta _{\perp e}$ with $\beta _{\perp \alpha
}=8\pi p{_{\perp \alpha }/B}_{0}^{2}$ where $p{_{\perp \alpha }}$ is the
perpendicular thermal pressure (similar definition for $\beta _{\Vert }$).
Furthermore, the parallel thermal velocity is defined as $v_{\Vert \alpha }=%
\sqrt{{2T_{\Vert \alpha }}/{m_{\alpha }}}$, and $\rho_{i}=({2{T_{\perp
i}/m_{p})^{1/2}/\Omega _{i}}}$ denotes the ion Larmor radius ($\Omega
_{i}=eB_{0}/m_{i}c$ is the ion gyrofrequency).

The growth rate given by Eq. (\ref{growth_rate}) has the same structure as
in the cold electron regime considered in the previous sections, and given first time
in \cite{hall} in the case of bi-Maxwellian ions and then generalized in \cite%
{PokhotelovSagdeev} and \cite{Hell07} to an arbitrary distribution function.
The first term within the curly brackets provides the threshold condition
which coincides with that given in \cite{Stix62},\cite{hall},\cite{Gary}. The second
one reflects the magnetic field line elasticity and the third one (where $F$
depends on the electron temperatures due to the coupling between the species
induced by the parallel electric field which is relevant for hot electrons)
provides the arrest of the instability at small scales by finite Larmor
radius (FLR) effects.

\subsection{Asymptotic model for hot electrons}

Now we extend to hot electrons the weakly nonlinear analysis developed for
cold electrons in the previous section. Since in this asymptotics, FLR
contributions appear only at the linear level, the idea is to use the drift
kinetic formalism to calculate the nonlinear terms. We show that the
equation governing the evolution of weakly nonlinear mirror modes has the
same form as in the case of cold electrons. In particular, the sign of the
nonlinear coupling coefficient that prescribes the shape of mirror
structures, is not changed, in the case of bi-Maxwellian distributions for both
electrons and ions, but can be changed for another distributions. This
equation is of gradient type  with a free energy (or a Lyapunov
functional) which is unbounded from below. This leads to finite-time
blowing-up solutions \cite{ZK12, KD}, associated with the existence of a
subcritical bifurcation \cite{KPS2007a,KPS2007b}. To describe subcritical
stationary mirror structures in the strongly nonlinear regime, we present an
anisotropic MHD model where the perpendicular and parallel pressures are
determined from the drift kinetic equations in the adiabatic approximation,
in the form of prescribed functions of the magnetic field amplitude only.

A main condition governing the nonlinear behavior of mirror modes is
provided by the force balance equation 
\begin{eqnarray}
&&-\nabla \left( p_{\perp }+\frac{B^{2}}{8\pi }\right) +\left[ 1+\frac{4\pi 
}{B^{2}}(p_{\perp }-p_{\Vert })\right] \frac{(\mathbf{B}\cdot \nabla )%
\mathbf{B}}{4\pi }  \nonumber \\
&&+\mathbf{B}(\mathbf{B}\cdot \nabla )\left( \frac{p_{\perp }-p_{\Vert }}{%
B^{2}}\right) -\nabla \cdot \mathbf{\Pi }=0,  \label{balance-nl1}
\end{eqnarray}%
where a gyroviscous contribution $\mbox{\boldmath $\Pi$}$ originating from
FLR effects (compare with (\ref{main})). Note that FLR contributions also
enter the gyrotropic pressures. Here the pressure tensor and its components
are viewed as the sum of the contributions of the various species. In
particular $p_{\perp }=\sum_{\alpha }p_{\perp \alpha }$ and $p_{\Vert
}=\sum_{\alpha }p_{\Vert \alpha }$. When concentrating on scales large
compared with the electron Larmor radius, the non-gyrotropic correction $%
\mbox{\boldmath $\Pi$} $ to the pressure tensor originates only from the
ions. As mentioned above, it is enough to retain this contribution only at
the linear level with respect to the amplitude of the perturbations. As in
the case of cold electrons, the other linear and nonlinear contributions can
be evaluated from the drift kinetic equation 
\begin{equation}
\frac{\partial f_{\alpha }}{\partial t}+v_{\Vert }\mathbf{b}\cdot \nabla
f_{\alpha }+\left( -\mu \mathbf{b}\cdot \nabla B+\frac{e_{\alpha }}{%
m_{\alpha }}E_{\Vert }\right) \frac{\partial f_{\alpha }}{\partial v_{\Vert }%
}=0  \label{mainkin1}
\end{equation}%
for each type of particles.

We ignore the transverse electric drift which is subdominant for mirror
modes. In this approximation, both ions and electrons move in the direction
of the magnetic field under the effect of the magnetic force $\mu \mathbf{\ b%
}\cdot \nabla B$ and the parallel electric field $E_{\Vert }=-\mathbf{b}%
\cdot \nabla \phi $ where the magnetic moment $\mu =v_{\perp }^{2}/(2B)$ is
an adiabatic invariant which plays the role of a parameter in Eq. (\ref%
{mainkin1}). Here $\phi $ is the electric potential. The quasi-neutrality
condition $n_{e}=n_{i}\equiv n$, where $n_{\alpha }=B\int f_{\alpha }d\mu
dv_{\Vert }d\varphi \equiv \int f_{\alpha }d^{3}v$, is used to close the
system and eliminate $E_{\Vert }$.

In this framework where FLR corrections are neglected, the gyrotropic
pressures $p_{\| \alpha }$ and $p_{\perp \alpha }$ are given in terms of the
corresponding distribution functions $f_{\alpha }$ by 
\begin{eqnarray}
&&p_{\alpha \Vert }=m_{\alpha }B\int v_{\Vert }^{2}f_{\alpha }d\mu dv_{\Vert
}d\varphi \equiv m_{\alpha }\int v_{\Vert }^{2}f_{\alpha }d^{3}v,  \nonumber
\\
&&p_{\alpha \perp }=m_{\alpha }B^{2}\int \mu f_{\alpha }d\mu dv_{\Vert
}d\varphi \equiv \frac{1}{2}m_{\alpha }\int v_{\perp }^{2}f_{\alpha }d^{3}v.
\nonumber
\end{eqnarray}

The equation governing the mirror dynamics is obtained perturbatively by
expanding Eqs. (\ref{balance-nl1}), (\ref{mainkin1}) and the
quasi-neutrality condition. In this approach, the pressure tensor elements
for each species are computed near a bi-Maxwellian equilibrium state
characterized by temperatures $T_{\perp\alpha }$ and $T_{\Vert\alpha }$ and
a constant ambient magnetic field $\mathbf{B_0}$ taken along the $z$%
-direction.

\subsection{Linear instability}

Before turning to the nonlinear regime, we briefly reformulate the linear
theory in the framework of the drift kinetic approximation, in order to
specify the notations.

From Eq. (\ref{balance-nl1}), linearized about the background field $\mathbf{%
\ B_{0}}$ by writing $\mathbf{B=B}_{0}+\widetilde{\mathbf{B}}$ ($B_{0}\gg 
\widetilde{B}$) with $\widetilde{\mathbf{B}}\sim e^{-i\omega t+i\mathbf{\
k\cdot r}}$, we arrive at Eq. (\ref{first}), where $p_{\perp }^{(1)}$ has to
be calculated from the linearized drift kinetic equation (\ref{mainkin1})
after elimination of the parallel electric field using the quasi-neutrality
condition. Note that as for the case of cold electrons, near the instability
threshold the leading terms in (\ref{first}) corresponding to perturbations
of perpendicular and magnetic pressures are compensated by each other and
therefore one needs to retain the next order terms responsible for both elasticity
of magnetic field lines and FLR corrections.

The linearized drift kinetic equation reads 
\begin{equation}
\frac{\partial f_{\alpha }^{(1)}}{\partial t}+v_{\Vert }\frac{\partial
f_{\alpha }^{(1)}}{\partial z}+\left( -\mu \frac{\partial \widetilde{B}_{z}}{%
\partial z}+\frac{e_{\alpha }}{m_{\alpha }}E_{\Vert }\right) \frac{\partial
f_{\alpha }^{(0)}}{\partial v_{\Vert }}=0,  \label{kin-1}
\end{equation}%
where we assume each $f_{\alpha }^{(0)}$ to be a bi-Maxwellian distribution
function 
\begin{equation}
f_{\alpha }^{(0)}=A_{\alpha }\exp \left[ -\frac{v_{\parallel }^{2}}{%
v_{\parallel \alpha }^{2}}-\frac{\mu B_{0}m_{\alpha }}{T_{\perp \alpha }}%
\right] ,  \label{bi-maxwell}
\end{equation}%
with $A_{\alpha }~=~n_{0}m_{\alpha }/(2\pi \sqrt{\pi }v_{\parallel \alpha
}T_{\perp \alpha })$.

In Fourier representation, Eq. (\ref{kin-1}) is solved as 
\begin{equation}
f_{\alpha }^{(1)}=-\frac{\mu \widetilde{B}_{z}+\frac{e_{\alpha }} {m_{\alpha
}}\phi }{\omega -k_{z}v_{\Vert }}k_{z}\frac{\partial f_{\alpha }^{(0)}}{
\partial v_{\Vert }}.  \label{kin-gen-1}
\end{equation}
The neutrality condition in the linear approximation reads 
\begin{equation}
\int f_{i}^{(1)}dv_{\parallel }d\mu d\varphi =\int
f_{e}^{(1)}dv_{\parallel}d\mu d\varphi,  \nonumber
\end{equation}
that allows one to express the potential $\phi$ in terms of $\widetilde{B}
_{z}$. We have 
\begin{equation}
\int f_{i}^{(1)}dv_{z}d\mu d\varphi =-\frac{n_{0}}{B_{0}T_{\parallel i}} %
\Big[ T_{\perp i}\frac{\widetilde{B}_{z}}{B_{0}}+e\phi \Big ] \Big [ 1+ 
\frac{i\sqrt{\pi }\omega }{|k_{z}|v_{\parallel i}}\Big ].  \nonumber
\end{equation}
Here we assume that $\omega /k_{z}\ll v_{{\Vert }i}=\sqrt{2T_{\Vert i}/m_{i}}
$, so that the contribution from the Landau pole is small ($\displaystyle{%
\xi ={\sqrt{\pi }\omega }/(|k_{z}|v_{\parallel i})\ll 1}$).

An analogous calculation for the electrons, neglecting the contribution of
the corresponding Landau resonance because of the small mass ratio (assuming
the ratio of the electron to ion temperatures is not too small), gives 
\begin{equation}
\int f_{e}^{(1)}dv_{z}d\mu d\varphi =-\frac{n_{0}}{B_{0}T_{\Vert e}}\left(
T_{\perp e}\frac{\widetilde{B}_{z}}{B_{0}}-e\phi \right) .  \nonumber
\end{equation}

The quasi-neutrality condition then reads 
\begin{equation}
\frac{1}{T_{\parallel i}}\Big[T_{\perp i}\frac{\widetilde{B}_{z}}{B_{0}}%
+e\phi \Big]\Big[1+\frac{i\sqrt{\pi }\omega }{|k_{z}|v_{\parallel i}}\Big]= 
\frac{1}{T_{\Vert e}}\Big[T_{\perp e}\frac{\widetilde{B}_{z}}{B_{0}}-e\phi %
\Big],  \nonumber
\end{equation}
and leads to the estimate 
\begin{equation}
e\phi \approx \frac{T_{\perp i}}{1+\theta _{\parallel }}\Big[(\theta _{\perp
}-\theta _{\parallel })-\frac{\theta _{\parallel }(1+\theta _{\perp })}{
1+\theta _{\parallel }}i\xi \Big]\frac{\widetilde{B}_{z}}{B_{0}}.
\label{phi-1}
\end{equation}
Thus, for cold electrons ($\theta_\perp = \theta_\| = 0$), $\phi$ vanishes
and the influence of the parallel electric field on the mirror instability
becomes negligible. Interestingly, when $\theta _{\perp }=\theta _{\parallel
}$, only the Landau pole contributes to 
\begin{equation}
e\phi \approx -\frac{T_{\perp i}\theta _{\parallel }}{1+\theta _{\parallel }}
i\xi \frac{\widetilde{B}_{z}}{B_{0}}.  \nonumber
\end{equation}

Now, it is necessary to evaluate 
\begin{equation}
p_{\perp }^{(1)}=2\frac{\widetilde{B}_{z}}{B_{0}}p_{\perp
}^{(0)}+B_{0}^{2}\sum_{\alpha }m_{\alpha }\int \mu f_{\alpha }^{(1)}d\mu
dv_{\Vert }d\varphi .  \nonumber
\end{equation}%
Using 
\begin{eqnarray}
\int \frac{k_{z}v_{\Vert }}{\omega -k_{z}v_{\Vert }}f_{i}^{(0)}d\mu
dv_{\Vert }d\varphi &=&-\frac{n_{0}}{B_{0}}(1+i\xi )  \nonumber \\
\int \frac{k_{z}v_{\Vert }}{\omega -k_{z}v_{\Vert }}f_{e}^{(0)}d\mu
dv_{\Vert }d\varphi &=&-\frac{n_{0}}{B_{0}},  \nonumber
\end{eqnarray}%
we get 
\begin{equation}
\sum_{\alpha }m_{\alpha }\int \mu f_{\alpha }^{(1)}d\mu dv_{\Vert }d\varphi
\approx -n_{0}\frac{T_{\perp i}^{2}}{T_{\parallel i}}\frac{\widetilde{B}_{z}%
}{B_{0}^{3}}\left( C+i\xi D\right) ,  \nonumber
\end{equation}%
where the coefficients $C$ and $D$, defined above, are both positive. In the
cold-electron limit, $C\rightarrow 2$ and $D\rightarrow 2$.

It is worth noting that the terms $\displaystyle{-\frac{\left( \theta
_{\perp }-\theta _{\parallel }\right) ^{2}}{\theta _{\parallel }(1+\theta
_{\parallel })}}$ in $C$ and $\displaystyle{\left( \theta _{\perp }-\theta
_{\parallel }\right) \frac{\left( 2+\theta _{\perp }+\theta _{\parallel
}\right) }{\left( 1+\theta _{\parallel }\right) ^{2}}}$ in $D$ originate
from the contributions of the electrostatic potential $\phi $ to $p_{\perp
}^{(1)}$, and vanish for $\theta _{\perp }=\theta _{\parallel }$.
Furthermore, in this limit, the real part of the perpendicular pressure
fluctuations is the sum of two independent contributions originating from
the ions and the electrons. Differently, only the ion Landau pole
contribution is retained in the imaginary part. We finally get 
\begin{equation}
p_{\perp }^{(1)}=\frac{\widetilde{B}_{z}}{B_{0}}n_{0}T_{\perp i}\left[
2(1+\theta _{\perp })-\frac{T_{\perp i}}{T_{\Vert i}}(C+i\xi D)\right] . 
\nonumber
\end{equation}%
Substituting this expression into the linearized force balance equation
yields 
\begin{eqnarray}
&&n_{0}T_{\perp i}\frac{T_{\perp i}}{T_{\parallel i}}Di\xi =2n_{0}T_{\perp
i}(1+\theta _{\perp })  \nonumber \\
&&\qquad \times \left[ 1-\frac{T_{\perp i}}{2T_{\parallel i}(1+\theta
_{\perp })}C+\frac{1}{\beta _{\perp }}+\frac{k_{z}^{2}}{k_{\perp }^{2}\beta
_{\perp }}\chi \right] ,  \nonumber
\end{eqnarray}%
and thus the linear growth rate 
\begin{eqnarray}
&&\gamma =|k_{z}|v_{\mathrm{th}{\Vert }i}\frac{2}{\sqrt{\pi }}\frac{%
T_{\parallel i}}{T_{\perp i}}\frac{1+\theta _{\perp }}{D}  \nonumber \\
&&\quad \times \left[ \frac{T_{\perp i}}{2T_{\parallel i}(1+\theta _{\perp })%
}C-1-\frac{1}{\beta _{\perp }}-\frac{k_{z}^{2}}{k_{\perp }^{2}\beta _{\perp }%
}\chi \right] ,  \label{gammanew}
\end{eqnarray}%
where $\chi =1+(\beta _{\perp }-\beta _{\Vert })/2$. It reproduces Eq. (\ref%
{growth_rate}) up to the FLR term which is not captured by the drift kinetic
approximation. 
As $\theta \rightarrow 0$, 
the growth rate reduces to the usual form given in \cite{VedenovSagdeev} 
\begin{equation}
\gamma =|k_{z}|v_{i{\Vert }}\frac{\beta _{\Vert }}{\sqrt{\pi }\beta _{\perp }%
}\left[ \frac{\beta _{\perp }}{\beta _{\Vert }}-1-\frac{1}{\beta _{\perp }}-%
\frac{k_{z}^{2}}{k_{\perp }^{2}\beta _{\perp }}\chi \right] .  \nonumber
\end{equation}%
In the presence of hot electrons, the mirror instability arises when 
\begin{eqnarray}
&&\Gamma =\frac{T_{\perp i}}{2T_{\parallel i}(1+\theta _{\perp })}\frac{1}{%
\theta _{\parallel }(\theta _{\parallel }+1)}\left[ (\theta _{\parallel
}+\theta _{\perp })^{2}+2\theta _{\parallel }(\theta _{\perp }^{2}+1)\right]
\nonumber \\
&&\qquad -1-\frac{1}{\beta _{\perp }}>0  \label{newthreshold1}
\end{eqnarray}%
and, near threshold, develops in quasi-perpendicular directions, making the
parallel magnetic perturbation dominant. This instability condition can be
also rewritten in the form given in \cite{Stix62}.

Note that the growth rate derived above is valid provided the condition $%
\gamma /k_{z}\ll v_{\mathrm{th}{\Vert i}}$ is fulfilled. Furthermore, the
instability is arrested by FLR effects at scales that are too small to be
captured by the drift kinetic asymptotics.

\subsection{General pressure estimates}

As demonstrated in Section 2 (see also \cite{KPS2007a,KPS2007b}), the
scalings (\ref{scaling}) resulting from the linear theory near threshold,
when $k_z$ and $k_{\perp}$ vary proportionally to $\varepsilon$ and $\sqrt{%
\varepsilon}$ respectively, while the instability growth rate behaves like $%
\sim \varepsilon^2$, imply an adiabaticity condition, or, in another words,
this leads to the stationary kinetic equation 
\begin{equation}
v_{\Vert }\mathbf{b}\cdot \nabla f_{\alpha }-(\mathbf{b}\cdot \nabla )\left[
\mu B+\frac{e_{\alpha }}{m_{\alpha }}\phi \right] \frac{\partial f_{\alpha }%
}{\partial v_{\Vert }}=0.  \label{Vlasov_stat}
\end{equation}%
It in fact turns out that Eq. (\ref{Vlasov_stat}) is exactly solvable, the
general solution being an arbitrary function of all integrals of motion $%
f_{\alpha }=g_{\alpha }(\mu ,W_{\alpha }, q)$ of the particle energy $%
\displaystyle{W_{\alpha }=\frac{v_{\Vert }^{2}}{2}+\mu B+\frac{e_{\alpha }}{%
m_{\alpha }}\phi }$, of $\mu $ and of variables $q$ responsible for labeling
the magnetic field lines. As we see in the previous case on the example of
cold electrons, the dependence on $q$ does not appear in the weakly
nonlinear regime, analyzed perturbatively. In the next section, we will
return to this question and discuss it in more detail. Below we will ignore
this dependence, considering only the case when $f_{\alpha }$ has two
arguments $\mu$ and $W_{\alpha }$.

To find the function $g_{\alpha }(\mu ,W_{\alpha })$ in this case, we use
the adiabaticity argument which means that, to leading order, $g_{\alpha }$
as a function of its arguments $\mu $ and $W_{\alpha }$ retains its form
during the evolution. Therefore, the function $g_{\alpha }(\mu ,W_{\alpha })$
is found by matching with the initial distribution function $f_{\alpha
}^{(0)}$, given by Eq. (\ref{bi-maxwell}) and corresponding to $\phi =0$ and $%
W_{\alpha }=\frac{v_{\Vert }^{2}}{2}+\mu B_{0}$. We get 
\begin{eqnarray}
&&g_{\alpha }(\mu ,W_{\alpha })=A_{\alpha }\exp \Big[-\frac{v_{\parallel
}^{2}}{v_{\parallel \alpha }^{2}}-\frac{\mu B_{0}m_{\alpha }}{T_{\perp
\alpha }}\Big]  \nonumber \\
\quad &=&A_{\alpha }\exp \Big[-\frac{2}{v_{\parallel \alpha }^{2}}\Big(\frac{%
v_{\Vert }^{2}}{2}+\mu B_{0}\Big)  \nonumber \\
&&\qquad +\mu B_{0}m_{\alpha }\Big(\frac{1}{T_{\parallel \alpha }}-\frac{1}{%
T_{\perp \alpha }}\Big)\Big]  \nonumber \\
&&\quad =A_{\alpha }\exp \Big [-\frac{2W_{\alpha }}{v_{\parallel \alpha }^{2}%
}+\mu B_{0}m_{\alpha }\Big(\frac{1}{T_{\parallel \alpha }}-\frac{1}{T_{\perp
\alpha }}\Big)\Big].  \label{g-alpha}
\end{eqnarray}%
Thus, $g_{\alpha }(\mu ,W_{\alpha })$ is a Boltzmann distribution function
with respect to $W_{\alpha }$ but, at fixed $W_{\alpha }$, it displays an
exponential growth relatively to $\mu $ if $T_{\perp \alpha }>$ $%
T_{\parallel \alpha }$. This effect can however be compensated by the
dependence of $W_{\alpha }$ in $\mu $. This means that only a fraction of
the phase space $(\mu ,W_{\alpha })$ is accessible, a property possibly
related with the concepts of trapped and untrapped particles.

Note that expanding Eq. (\ref{g-alpha}) relatively to $\widetilde{B}_{z}$
and $\phi ^{(1)}$ reproduces the first order contribution to the
distribution function (\ref{kin-gen-1}) with $\omega=0$ and also the
corresponding expression for the second order correction (\ref{second-order}%
) found in the previous section (see also \cite{KPS2007a}, \cite{KPS2007b})
in the case of cold electrons. It should be emphasized that Eq. (\ref%
{g-alpha}) only assumes adiabaticity and remains valid for finite
perturbations.

The function $g_{\alpha }$ can also be rewritten in terms of $v_{\Vert }$, $%
v_{\perp }$ and $\phi $ as 
\begin{eqnarray}
&&g_{\alpha }=A_{\alpha }\exp \Big[-\frac{m_{\alpha }v_{\Vert }^{2}}{
2T_{\parallel \alpha }}-\frac{e_{\alpha }\phi }{T_{\parallel \alpha }}\Big] %
\times  \nonumber \\
&&\qquad \exp \left\{ -\frac{m_{\alpha }v_{\perp }^{2}}{2T_{\perp \alpha }} %
\Big (\frac{T_{\perp \alpha }}{T_{\parallel \alpha }}-\frac{B_{0}}{B}\Big [ 
\frac{T_{\perp \alpha }}{T_{\parallel \alpha }}-1\Big ]\Big )\right\} , 
\nonumber
\end{eqnarray}%
which can be viewed as the bi-Maxwellian distribution function with the
renormalized transverse temperature 
\begin{equation}
T_{\perp \alpha }^{(eff)}=T_{\perp \alpha }\left[ \frac{T_{\perp \alpha }}{
T_{\parallel \alpha }}-\frac{B_{0}}{B}\Big(\frac{T_{\perp \alpha }}{
T_{\parallel \alpha }}-1\Big)\right] ^{-1}.  \nonumber
\end{equation}%
Note the Boltzmann factor $\exp {-[e_{\alpha }\phi /T_{\parallel \alpha }]}$
in the expression of $g_{\alpha }$. For cold electrons, the ion distribution
function was obtained in \cite{Const02} by assuming that the distribution
remains bi-Maxwellian and owing to the invariance of the kinetic energy and
of the magnetic moment. This estimate obtained by neglecting both time
dependency (and consequently the Landau resonance) and finite Larmor radius
corrections reproduces the closure condition given in \cite{PRS06}.

After rewriting Eq. (\ref{g-alpha}) in the form 
\begin{equation}
g_{\alpha }=A_{\alpha }\exp \Big[-\frac{e_{\alpha }\phi }{T_{\parallel
\alpha }}-\frac{v_{\Vert }^{2}}{v_{\parallel \alpha }^{2}}-\frac{\mu
B_{0}m_{\alpha }}{T_{\perp \alpha }}\Big(1+\frac{T_{\perp \alpha }}{
T_{\parallel \alpha }}\frac{B-B_{0}}{B_{0}}\Big)\Big],  \nonumber
\end{equation}
the quasi-neutrality condition gives 
\begin{eqnarray}
&&\left( 1+\frac{T_{\perp i}}{T_{\parallel i}}\frac{B-B_{0}}{B_{0}}\right)
^{-1}\exp \left( -\frac{e\phi }{T_{\parallel i}}\right) =  \nonumber \\
&&\left( 1+\frac{T_{\perp e}}{T_{\parallel e}}\frac{B-B_{0}}{B_{0}}\right)
^{-1}\exp \left( \frac{e\phi }{T_{\parallel e}}\right)  \nonumber
\end{eqnarray}
or 
\begin{eqnarray}
&&e\phi =(T_{\parallel i}^{-1}+T_{\parallel e}^{-1})^{-1}\times  \nonumber \\
&&\log \left[ \left( 1+\frac{T_{\perp e}}{T_{\parallel e}}\frac{B-B_{0}}{
B_{0}}\right) \left( 1+\frac{T_{\perp i}}{T_{\parallel i}}\frac{B-B_{0}}{
B_{0}}\right) ^{-1}\right] .  \label{potential}
\end{eqnarray}%
Interestingly, the electron density 
\begin{equation}
n_{e}=n_{0}\frac{B}{B_{0}}\left( 1+\frac{T_{\perp e}}{T_{\parallel e}}\frac{
B-B_{0}}{B_{0}}\right) ^{-1}\exp \left[ \frac{e\phi }{T_{\parallel e}}\right]
\nonumber
\end{equation}%
has the usual Boltzmann factor $\exp \left[ e\phi /T_{\parallel e}\right] $
and also an algebraic prefactor depending on the magnetic field $B$. In the
case of isotropic electron temperature ($T_{\perp e}=T_{\parallel e}\equiv
T_{e}$), the electron density has the usual Boltzmann form $n_{e}=n_{0}\exp %
\left[ e\phi /T_{e}\right] $.

The above formula for $\phi $ shows that the potential vanishes in two
cases: for cold electrons and when electron and ion temperature anisotropies 
$a_e$ and $a_i$ are equal, a case first time mentioned in the linear theory
of the mirror instability \cite{Stix62,hasegawa,hall}.

Equation (\ref{potential}) allows one to evaluate explicitly the
perpendicular pressure for each species 
\begin{eqnarray}
&&p_{\perp \alpha }=m_{\alpha }B^{2}\int \mu g_{\alpha }d\mu dv_{\Vert
}d\varphi  \nonumber \\
&&\ \ =n_{0}T_{\perp \alpha }\frac{B^{2}}{B_{0}^{2}}\Big(1+\frac{T_{\perp
\alpha }}{T_{\parallel \alpha }}\frac{B-B_{0}}{B_{0}}\Big)^{-2}\exp \Big(-%
\frac{e_{\alpha }\phi }{T_{\parallel \alpha }}\Big),  \nonumber
\end{eqnarray}%
where $e\phi $ is given by Eq. (\ref{potential}).

Hence, simple algebraic procedure gives the following expression for the
parallel pressure \cite{KPS2012a}, \cite{KPRS2014}:%
\begin{equation}
p_{\Vert }=n_{0}(T_{\Vert i}+T_{\Vert e})\frac{1+u}{\left( 1+a_{e}u\right)
^{c_{e}}\left( 1+a_{i}u\right) ^{c_{i}}},  \label{total-parallel}
\end{equation}%
where $u=B/B_{0}-1$, $a_{\alpha }=T_{\perp \alpha }/T_{\parallel \alpha }$
is the parameter characterizing the anisotropy of distribution function $%
f_{\alpha }$, and $c_{\alpha }=T_{\parallel \alpha }(T_{\parallel
e}+T_{\parallel i})^{-1}$ in the case of a proton-electron plasma. As it
will be shown in the next section, the perpendicular pressure can be easily
found by means of the general relation%
\begin{equation}
p_{\perp }=p_{\Vert }-B\frac{dp_{\Vert }}{dB}.  \label{total-perp}
\end{equation}%
Substitution of (\ref{total-parallel}) into this expression yeilds 
\begin{equation}
p_{\perp }=p_{\Vert }(1+u)\left( \frac{c_{e}a_{e}}{1+a_{e}u}+\frac{c_{i}a_{i}%
}{1+a_{i}u}\right) .  \label{total-perp1}
\end{equation}%
Hence one can see that both pressures have the singularities at $%
u=-a_{\alpha }^{-1}$ corresponding to the magnetic field 
\begin{equation}
B_{s}=B_{0}\frac{a_{\alpha }-1}{a_{\alpha }}<B_{0}.  \label{singularB}
\end{equation}%
In the limiting case of cold electrons, $p_{\Vert }=n_{0}T_{\Vert
}(1+u)(1+au)^{-1}$ displays a pole singularity. Here, $T_{\Vert }$ and the
anisotropy parameter $a$ correspond to ions only. Such an equation of state
was previously derived by a quasi-normal closure of the fluid hierarchy \cite%
{PRS06}.

The above singularities are presumably related to an overestimated
contribution from large $\mu $, corresponding either to small $B$ or to
large a transverse kinetic energy. In both cases, the applicability of the
drift approximation breaks down and we are thus led to introduce some
cut-off type correction near $\mu _{\alpha }^{\ast }$. In a simple variant,
we take $f_{\alpha }=\tilde{C}_{\alpha }\exp (-m_{\alpha }W_{\alpha
}/T_{\parallel \alpha })$ at $\mu >\mu _{\alpha }^{\ast }$, with some
positive constant $\tilde{C}_{\alpha }$, and $f_{\alpha }$ retains its
original form (\ref{g-alpha}) for $\mu \leq \mu _{\alpha }^{\ast }$. For
cold electrons, the parallel ion pressure is modified into $p_{\Vert
}=n_{0}T_{\Vert }G(B,r)$ with 
\[
G(B,r)=\frac{1}{1+C}\left[ \frac{(B_{0}-B_{s})B}{B_{0}(B-B_{s})}%
R(B,r)+Ce^{r(B_{0}-B)}\right] , 
\]%
and 
\[
R(B,r)=\frac{\exp [-r(B-B_{s})]-1}{\exp [-r(B_{0}-B_{s})]-1}. 
\]%
Here, $C$ is a (small) constant, and $r=m\mu ^{\ast }/T_{\Vert }$.
Noticeably, regularization leads to a non-singular positive pressure for all 
$B$, including when $B\rightarrow 0$. The modification for $p_{\Vert }$ in
the case of hot electrons is not specified here because the expressions are
algebraically much more cumbersome but do not involve any additional
difficulty.

After these remarks, one can easily derive the asymptotic model with account
of hot electrons. The basic idea is the same as we used already while
derivation the model (\ref{main}) for cold electrons. To derive the
asymptotic model, we can of course forget about renormalization of the
function $G(B,r)$ because we need to consider the expansion of $p_{\perp }$
with respect to small amplitude $u $ by taking into account in this
expansion only the second term $\sim u^{2}$ which defines the nonlinear
coupling coefficient for (\ref{main}). For (\ref{total-perp1}) the quadratic
contributions originating from $p_{\perp }^{(2)}+\left( B-B_{0}\right)
^{2}/(8\pi )$ are collected in a term $\Lambda \left( \frac{B-B_{0}}{B_{0}}%
\right) ^{2}$ with 
\begin{eqnarray}
\Lambda &=&n_{0}\Big \{T_{\perp i}\Big (3a_{i}^{2}-4a_{i}+1  \nonumber \\
&&+c_{i}(a_{e}-a_{i})\Big [\frac{1}{2}(1+c_{i})(a_{e}-a_{i})-2+3a_{i}\Big ]%
\Big )  \nonumber \\
&&+T_{\perp e}\Big (3a_{e}^{2}-4a_{e}+1+c_{e}(a_{e}-a_{i})  \nonumber \\
&&\times \Big [\frac{1}{2}(1+c_{e})(a_{e}-a_{i})+2-3a_{e}\Big ]\Big )\Big \}+%
\frac{B_{0}^{2}}{8\pi }.  \label{eqlambda}
\end{eqnarray}%
The value $\Lambda _{c}$ of $\Lambda $ at threshold is obtained by
expressing $\displaystyle{\frac{B_{0}^{2}}{8\pi }}$ by means of Eq. (\ref%
{newthreshold}), which gives 
\begin{eqnarray}
\Lambda _{c} &=&n_{0}\Big \{T_{\perp i}\Big [3a_{i}^{2}-4a_{i}+1  \nonumber
\\
&&+c_{i}(a_{e}-a_{i})\Big (\frac{1}{2}(1+c_{i})(a_{e}-a_{i})-2+3a_{i}\Big ) 
\nonumber \\
&&-\frac{1}{2}\Big (2-2a_{i}-c_{i}(a_{e}-a_{i})\Big )\Big ]  \nonumber \\
&&+T_{\perp e}\Big [3a_{e}^{2}-4a_{e}+1+c_{e}(a_{e}-a_{i})  \nonumber \\
&&\times \Big (\frac{1}{2}(1+c_{e})(a_{e}-a_{i})+2-3a_{e}]  \nonumber \\
&&-\frac{1}{2}\Big (2-2a_{e}+c_{e}(a_{e}-a_{i})\Big )\Big ]\Big \}. 
\nonumber
\end{eqnarray}%
After some algebra, one gets 
\begin{eqnarray}
&&\frac{\lambda _{c}}{\alpha _{i}}=\frac{T_{\perp i}}{T_{\parallel i}}\Big [%
3+3\frac{\theta _{\perp }^{3}}{\theta _{\parallel }^{2}}-\frac{1}{2}\frac{%
\left( \theta _{\perp }-\theta _{\parallel }\right) ^{2}}{\theta _{\parallel
}^{2}\left( 1+\theta _{\parallel }\right) ^{2}}  \nonumber \\
&&\qquad \times \left( 4\theta _{\perp }+4\theta _{\parallel
}^{2}+\allowbreak 5\left( \theta _{\perp }+1\right) \theta _{\parallel
}\right) \Big]  \nonumber \\
&&\qquad -\frac{3}{2\theta _{\parallel }\left( 1+\theta _{\parallel }\right) 
}\left[ \left( \theta _{\perp }+\theta _{\parallel }\right) ^{2}+2\theta
_{\parallel }(1+\theta _{\perp }^{2})\right] ,  \label{eqlambda_c}
\end{eqnarray}%
where $\lambda _{c}=\Lambda _{c}/(n_{0}T_{\perp i})$.

Supplementing the corresponding quadratic terms in Eq. (\ref{growth_rate})
leads, at the order of the expansion, to the dynamical equation 
\begin{eqnarray}
&&\frac{\partial}{\partial t}\frac{\widetilde{B}_{z}}{B_{0}}=\frac{2}{\sqrt{%
\pi }}\frac{T_{\Vert i}}{T_{\perp i}}\frac{v_{\mathrm{th}\Vert i}}{D}\left( {%
-\mathcal{H}}\partial _{z}\right)  \nonumber \\
&&\quad \times \Big \{\Big [\frac{T_{\perp i}}{T_{\Vert i}}\frac{C}{2}%
-(1+\theta _{\perp })\Big (1+\frac{1}{\beta _{\perp }}\Big )\Big ]\frac{%
\widetilde{B}_{z}}{B_{0}}  \nonumber \\
&&\quad -(1+\theta _{\perp })\frac{1}{\beta _{\perp }}\Big (1+\frac{\beta
_{\perp }-\beta _{\Vert }}{2}\Big )(\Delta )^{-1}\partial _{zz}\frac{%
\widetilde{B}_{z}}{B_{0}}  \nonumber \\
&&\quad +\frac{3}{4}\Big (\frac{T_{\perp i}}{T_{\Vert i}}-1\Big )%
(1+F)r_{L}^{2}\Delta _{\perp }\frac{\widetilde{B}_{z}}{B_{0}}-\frac{\lambda
_{c}}{2}\Big (\frac{\widetilde{B}_{z}}{B_{0}}\Big )^{2}\Big \}  \nonumber
\end{eqnarray}%
that extends the result of \cite{KPS2007a, Calif08} valid for cold
electrons. As demonstrated in \cite{KPS2007a,KPS2007b}, the sign of the
nonlinear coupling $\lambda _{c}$ defines the type of subcritical
structures, namely holes ($\lambda _{c}>0$) or humps ($\lambda _{c}<0$). It
turns out that the sign of the nonlinear coupling can be determined
analytically in a few special cases.

\noindent \textit{(i) Limit $\theta_\| \ll \theta_\perp$:} 
\begin{equation}
\frac{\Lambda _{c}}{n_{0}T_{\perp i}a _{i}}=\frac{\theta _{\perp }^{2}}{%
\theta _{\parallel }}\left( \frac{T_{\perp e}}{T_{\parallel e}}-\frac{3}{2}%
\right) >0.  \nonumber
\end{equation}

\noindent \textit{(ii) Equal anisotropies ($\theta
_{\perp}=\theta_{\parallel }$)} 
\begin{eqnarray}
\Lambda _{c} &=&n_{0}(T_{\perp i}+T_{\perp e})\left( 3a ^{2}-4a +1\right) 
\nonumber \\
&&-n_{0}(T_{\perp i}+T_{\perp e})\left( 1-a \right) =3a \frac{B_{0}^{2}}{%
8\pi }>0.  \nonumber
\end{eqnarray}

\noindent \textit{(iii) Isotropic electron temperature:} The coefficient $%
\Lambda _{c}$ can be rewritten in the form 
\begin{eqnarray}
\Lambda _{c} &=&n_{0}(a _{i}-1)\{T_{\perp i}\Big((3a _{i}-1)  \nonumber \\
&&+c_{i}\Big[\frac{1}{2}(1+c_{i})\left( \alpha _{i}-1\right) +2-3a _{i}\Big]%
\Big )  \nonumber \\
&&+T_{e}c_{e}\Big[\frac{1}{2}\left( 1+c_{e}\right) \left( a _{i}-1\right) +1%
\Big]\}+\frac{B_{0}^{2}}{8\pi }.  \nonumber
\end{eqnarray}
Furthermore, at threshold 
\begin{equation}
\frac{1}{2}n_{0}(a _{i}-1)\left[ T_{\perp i}\left( 2-c_{i}\right) +T_{\perp
e}c_{e}\right] =\frac{B_{0}^{2}}{8\pi }>0.  \nonumber
\end{equation}%
Hence, we simultaneously have two inequalities $a _{i}>1$ and $T_{\perp
e}c_{e}>T_{\perp i}(c_{i}-2)$. Therefore, 
\begin{eqnarray}
\Lambda _{c} &=&n_{0}(a _{i}-1)\Big \{T_{\perp i}\Big((3a _{i}-1)  \nonumber
\\
&&+c_{i}\Big [\frac{1}{2}(1+c_{i})\Big(a _{i}-1\Big)+2-3a _{i}\Big]\Big) 
\nonumber \\
&&+T_{e}c_{e}\Big[\frac{1}{2}(1+c_{e})(a _{i}-1)+1\Big ]\}  \nonumber \\
&&+\frac{1}{2}n_{0}(a _{i}-1)\left[ T_{\perp i}\left( 2-c_{i}\right)
+T_{\perp e}c_{e}\right]  \nonumber \\
&=&n_{0}(a _{i}-1)\Big \{T_{\perp i}\Big(3a _{i}(1-c_{i})  \nonumber \\
&&+c_{i}\Big[\frac{1}{2}(1+c_{i})\left( a _{i}-1\right) +\frac{3}{2}\Big]%
\Big)  \nonumber \\
&&+T_{e}c_{e}\Big[2+\frac{1}{2}\left( 1+c_{e}\right) \left( a _{i}-1\right) %
\Big]\Big \},  \nonumber
\end{eqnarray}%
which is positive because $\displaystyle{1-c_{i}\equiv c_{e}=\frac{1}{%
1+\theta _{\parallel }}>0}$ and $a _{i}-1>0$.

\begin{figure}[t]
\centerline{
\includegraphics[width=0.5\textwidth]{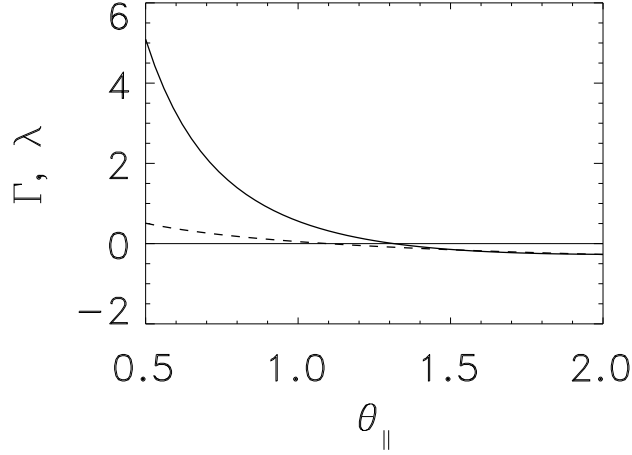}
}
\caption{Variation with $\protect\theta_\|$ of the distance to
threshold $\Gamma$ given by Eq. (\protect\ref{newthreshold1}) (dashed line)
and of the normalized nonlinear coupling coefficient $\protect\lambda$
(solid line) evaluated from Eq. (\protect\ref{eqlambda}) for $\protect\theta%
_{\perp}=1$ , $a_i=1.1$ and $\protect\beta_{\perp i} = 10$.}
\label{fig1}
\end{figure}

\begin{figure}[t]
\centerline{
\includegraphics[width=0.5\textwidth]{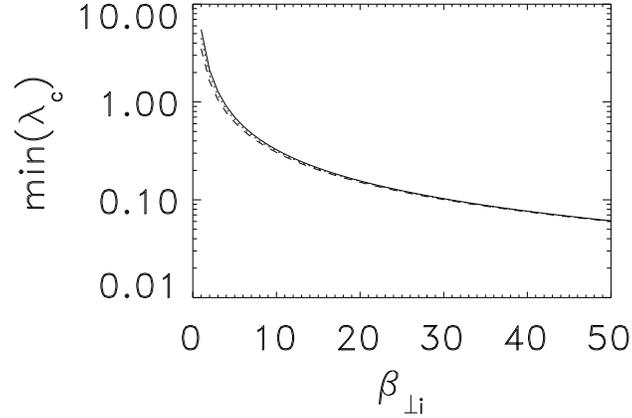}
}
\caption{Variation with $\protect\beta_{\perp i}$ of the minimum $%
\mathrm{min}\,( \protect\lambda_c)$ of the normalized nonlinear coupling
coefficient taken in an interval of values of $a_p$ between $0$ and $a_{p1}(%
\protect\beta_{\perp i})$, defined such that the threshold is obtained for a
value of $\protect\theta_\|$ equal to $100$, for $\protect\theta_\perp= 0.2$
(solid line), $\protect\theta_\perp=1$ (dotted line) and $\protect\theta%
_\perp=5$ (dashed line).}
\label{fig2}
\end{figure}

\noindent \textit{(iv) More general conditions:} A numerical approach was
used. For this purpose it is of interest to display in Fig. 1, for typical
values of the parameters taken here as $\theta _{\perp }=1$, $a _{i}=1.1$
and $\beta _{\perp i}=10$, the distance to threshold $\Gamma $ (dashed line)
given by Eq. (\ref{newthreshold1}) and the non-dimensional nonlinear
coupling coefficient $\lambda =\Lambda /(n_{0}T_{\perp i})$ (solid line),
with $\Lambda $ given by Eq. (\ref{eqlambda}), as a function of $\theta
_{\Vert }$. This graph is typical of the general behavior of these functions
and shows that they are both decreasing as $\theta _{\Vert }$ increases,
with $\lambda $ possibly reaching negative values, but only below threshold.
In order to show that the value $\lambda _{c}$, given by Eq. (\ref%
{eqlambda_c}), of $\lambda $ at threshold is positive in a wider range of
parameters, we display in Fig. 2, as a function of $\beta _{\perp i}$ for $%
\theta _{\perp }=0.2$ (solid line), $\theta _{\perp }=1$ (dotted line) and $%
\theta _{\perp }=5$ (dashed line), the quantity $\mathrm{{min}\,(\lambda
_{c})}$ obtained after minimizing $\lambda _{c}$ in an interval of values of 
$a_{p}$ between $0$ and $a_{p1}(\beta _{\perp i})$. The latter quantity is
arbitrarily defined such that the threshold is obtained for a value of $%
\theta _{\Vert }$ equal to $100$. This graph shows that $\mathrm{min}%
(\lambda _{c})$ varies little with $\theta _{\perp }$ but is very sensitive
to $\beta _{\perp i}$. As the latter parameter is increased, $\mathrm{{min}%
\,(\lambda _{c})}$ decreases towards zero but remains always positive.
Although this numerical observation is definitively not a rigorous proof, it
convincingly shows that $\Lambda $ should remain positive in the parameter
regime of physical interest.

Thus, we can see that in the case of the bi-Maxwelian distribution functions for both
ions and electrons (i) the asymptotic model has the same structure as in the
cold case, and (ii) it predicts the formation of magnetic holes which is
defined by the sign of the coupling coefficient $\Lambda $ . If the
disributions are different for the bi-Maxwellian ones we can expect change
of the  sign of $\Lambda $and appearance of magnetic structures in the form of
humps respectively.  In the next sections, we show how such mirror
structures can be found for arbitrary distributions for both \ electrons and
ions based on the variational principle when both pressures are functions of
the magnetic field amplitude only. \ 

\section{Variational principle for stationary anisotropic MHD}

\bigskip

As we saw in the previous sections, the nonlinearity for mirror modes
originates from  equations (\ref{balance-nl1}) which represent the
anisotropic MHD in a static regime supplemented by corrections due to the FLR\ effects.
Secondly, another origin of the nonlinearity comes from the drift kinetic
equations, in particular, for the asymptotic model (\ref{main}) it comes
from the stationary kinetic equations (\ref{Vlasov_stat}). Thus, the static
anisotropic MHD together with the stationary drift kinetic equations describe
the nonlinear development of the mirror modes and its possible saturation in
the form of static structures. In this section, we give \ formulation of the
variational principle for such structures and establish connection it with
the free energy formalism developed for the asymptotic model (\ref{main}).

\subsection{Gyrotropic pressure balance}
 We start from the pressure
balance equation for a static gyrotropic MHD equilibrium%
\begin{equation}
0=-\nabla \cdot \mathbf{P}+\frac{1}{c}\left[ \mathbf{j}\times \mathbf{B}%
\right] ,  \label{MHD}
\end{equation}%
where the current $\mathbf{j}$ is defined from the Maxwell equation as%
\textbf{\ }$\mathbf{j}{=\frac{c}{4\pi }\nabla \times }\mathbf{B}$, and the
pressure tensor $\mathbf{P}$ is assumed to be gyrotropic. The solvability
conditions read $\mathbf{B}\cdot (\nabla \cdot \mathbf{P})\mathbf{=}0\mathbf{%
,}$ and $\mathbf{j}\cdot (\nabla \cdot \mathbf{P})\mathbf{=}0$.

In terms of the tension tensor $S_{ij}=\Pi _{\perp }\left( \delta
_{ij}-b_{i}b_{j}\right) +S_{\parallel }b_{i}b_{j}$, Eq. (\ref{MHD}) takes
the divergence form $\frac{\partial }{\partial x_{j}}\Pi _{ij}=0$ where $S
_{\perp }=p_{\perp }+B^{2}/(8\pi )$ and $S_{\parallel }=p_{\parallel
}-B^{2}/(8\pi )$, and the perpendicular and parallel pressures $p_{\perp
}=\Sigma _{\alpha }p_{\perp \alpha }$ and $p_{\perp }=\Sigma _{\alpha
}p_{\perp \alpha }$ are the sum of the contributions of the various particle
species $\alpha $. They are expressed as $p_{\perp \alpha }=m_{\alpha
}B^{2}\int \mu f_{\alpha }dv_{\parallel }d\mu $ and $p_{\parallel \alpha
}=m_{\alpha }B\int v_{\parallel }^{2}f_{\alpha }dv_{\parallel }d\mu $, in
terms of the distribution functions $f_{\alpha }$, which satisfy the
stationary drift kinetic equations 
\begin{equation}
v_{\parallel }\nabla _{\parallel }f_{\alpha }-\left[ \mu \nabla _{\parallel
}B+\frac{e_{\alpha }}{m_{\alpha }}\nabla _{\parallel }\phi \right] \frac{%
\partial f_{\alpha }}{\partial v_{\parallel }}=0.  \label{kineticEqs}
\end{equation}%
These equations are supplemented by the quasi-neutrality condition 
\begin{equation}
\sum_{\alpha }e_{\alpha }B\int f_{\alpha }dv_{\parallel }d\mu =0,
\label{neutral}
\end{equation}
that allows one to eliminate the electric potential.

We consider partial solutions of the stationary kinetic equations (\ref%
{kineticEqs}) which are expressed in terms of two integrals of motion: the
energy of the particles $W_{\alpha }={v_{\parallel }^{2}}/{2}+\mu
B+(e_{\alpha }/m_{\alpha })\phi $ and their magnetic moment $\mu $. In
general, the solution can also depend on integrals which label the magnetic
field lines \cite{Grad1967}. The choice $f_{\alpha }=f_{\alpha }(W_{\alpha
},\mu )$, as it will be shown further, can be matched with the solution
found perturbatively for weakly nonlinear mirror modes within the asymptotic
model (\ref{main}) . In this case, the parallel and perpendicular pressures
for the individual species and also the total pressures are functions of $B$
only. We write $p_{\perp \alpha }=p_{\perp \alpha }(B)$ and $p_{\parallel
\alpha }=p_{\parallel \alpha }(B)$. As seen in the next subsection, this
property plays a very central role in the forthcoming analysis.

\subsection{Identity in the parallel direction and variational principle} 
According to
Ref. \cite{NorthropWhiteman1964}, the anisotropic pressure balance equation (%
\ref{MHD})can be easily reformulated as follows:%
\begin{equation}
-\nabla p_{\parallel }-\frac{1}{B}(p_{\perp }-p_{\parallel })\nabla B=\left[ 
\mathbf{B}\times \left[ \nabla \times \left( \frac{p_{\perp }-p_{\parallel }%
}{B^{2}}+\frac{1}{4\pi }\right) \mathbf{B}\right] \right] .  \label{MHD1}
\end{equation}
Hence projection along the magnetic field gives 
\begin{equation}
-\nabla _{\parallel }p_{\Vert }-\frac{4\pi \left( p_{\perp }-p_{\Vert
}\right) }{B^{2}}\nabla _{\parallel }\frac{B^{2}}{8\pi }=0,
\label{parallel-projection}
\end{equation}%
which coincides with Eq. (9.2) of Shafranov's review \cite{shafranov-66}. It
is possible to prove that Eq. (\ref{parallel-projection}), being solvability
condition to (\ref{MHD}), reduces to an identity, by means of the
statioinary kinetic equations (\ref{kineticEqs}) \ together with the
quasi-neutrality condition (\ref{neutral}). To our knowledge, first time
this fact was established by J.B. Taylor \cite{Taylor1963, Taylor1964} \ and
later by many others (see, for instance, \cite{NorthropWhiteman1964,
Grad1967, HallMcNamara1975, ZakharovShafranov}). Since the pressures depend
on $B$ only, Eq. (\ref{parallel-projection}) reduces to 
\begin{equation}
-\frac{dp_{\Vert }}{dB}=\frac{\left( p_{\perp }-p_{\Vert }\right) }{B}.
\label{identity3}
\end{equation}%
In this partial case the system (\ref{MHD1}) is written as

\begin{equation}
\left[ \mathbf{B}\times \left[ \nabla \times \left( \frac{p_{\perp
}-p_{\parallel }}{B^{2}}+\frac{1}{4\pi }\right) \mathbf{B}\right] \right] =0.
\label{MHD2}
\end{equation}

The existence of the identity (\ref{parallel-projection}) and its partial
formulation (\ref{identity3}) means that for stationary states, the pressure
balance provides only two scalar equations which, together with the
condition $\nabla \cdot \mathbf{B}=0$, leads to a closed system of three
equations for the three magnetic field components.

It can be easilty shown that Eq. (\ref{MHD2}) can be written in the
following variational form:%
\begin{equation}
\left[ \mathbf{B}\times \frac{\delta \mathcal{F}}{\delta \mathbf{A}}\right]
=0
\end{equation}%
where $\mathcal{F}$ is given by the expression 
\[
\mathcal{F}\mathcal{=}\int [B^{2}/(8\pi )-p_{\parallel }(B)]d^{3}\mathbf{r}, 
\]
and $\mathbf{A}$ is the vector potential: $\mathbf{B=}\left[ \nabla \times 
\mathbf{A}\right] .$ In the pure 2D geometry with $\mathbf{B=(}%
B_{x},B_{y},0) $ when the vector potential $\mathbf{A}$ has only one
non-zeroth component $\psi $ ($z$-component) for description of stationary
state we arrive at the variational principle $\delta \mathcal{F}\mathcal{=}0$%
, formulated in our paper \cite{KPRS2014}. It is evident also that we have
the same variational principle for stationary structures in $r-z$ geometry
when $\mathbf{B=(}B_{r},B_{z},0).$

\bigskip Note, that Eq. (\ref{MHD1}) can be written also as 
\begin{equation}
\left[ \nabla \times \left( 1+4\pi \frac{p_{\perp }-p_{\parallel }}{B^{2}}%
\right) \mathbf{B}\right] =\chi \mathbf{B}  \label{chi}
\end{equation}%
where for scalar function $\chi $ we have the equation$\ (\mathbf{B\cdot
\nabla }\chi )=0$. This equation shows that $\chi $ is constant along each
magnetic line. If the line is not closed so that at $r\rightarrow \infty $ $%
\mathbf{B}$ tends to the constant magnetic field $\mathbf{B}_{0}$ and
besides there both pressures $p_{\perp }$ and $p_{\parallel }$are also
constant then for all such lines $\chi =0.$ Indeed, as we will show below,
the equation for the stationary structures following from the guiding center
formalism coincides with Eq. (\ref{chi}) at $\chi =0.$

\subsection{Derivation of the varitional principle from the guiding-center
formalism}

\bigskip

To derive the variational principle for stationary mirror structures a
three-dimensional, previously established for 2D configurations \cite%
{KPRS2014}, we now employ the Hamiltonian theory of guiding-center motion as
stated in Section III of Ref.\cite{RMP2009}.

Let us first consider a sort of particles with mass $m$ and electric charge $%
e$. Instead of the particle position ${\mathbf{x}}$ and velocity ${\mathbf{v}%
}$, we introduce new coordinates in the phase space: position ${\mathbf{X}}$
of the guiding center, parallel velocity component $u$ along the magnetic
field $\mathbf{B}({\mathbf{X}},t)$, magnetic moment $\mu \approx m|{\mathbf{v%
}}_{\perp }|^{2}/2B(X,t)$, and a gyroangle $\zeta $. The dynamics of these
new unknown functions is determined by the following approximate Lagrangian
(see derivation in Ref.\cite{RMP2009}), valid to the lowest order on spatial
derivatives, 
\begin{equation}
L({\mathbf{X}},u,\mu ,\zeta )\approx \left[ \frac{e}{c}{\mathbf{A}}({\mathbf{%
X}},t)+mu{\mathbf{b}}({\mathbf{X}},t)\right] \dot{\mathbf{X}}+\frac{mc}{e}%
\mu \dot{\zeta}-\frac{m}{2}u^{2}-\mu B({\mathbf{X}},t)-e\phi ({\mathbf{X}}%
,t),  \label{Lagr_mu}
\end{equation}%
where ${\mathbf{A}}({\mathbf{X}},t)$ is the vector electromagnetic
potential, $\phi ({\mathbf{X}},t)$ is the scalar electric potential, $%
\mathbf{b}=\mathbf{B}/B$ is the unit tangent vector. We see that $\zeta $ is
a cyclical variable in this (adiabatic) approximation, and therefore $\mu $
is a nearly conserved quantity.

It will be important that the volume element in the non-canonical phase
space $(\mathbf{X},u,\mu ,\zeta )$ contains a non-constant Jacobian $J$, 
\[
d\mathcal{V}=d{\mathbf{x}}d{\mathbf{v}}=J(\mathbf{X},u)d^{3}\mathbf{X}dud\mu
d\zeta \propto \lbrack B+\frac{mc}{e}u(\mathbf{b}\cdot \mbox{curl}\,\mathbf{b%
})]d^{3}\mathbf{X}dud\mu d\zeta /(2\pi ). 
\]%
Another important formula determines velocity of guiding center in
stationary fields: 
\begin{equation}
\dot{\mathbf{X}}\approx u\mathbf{b}+\frac{mcu^{2}}{eB}[\mbox{curl}\,\mathbf{b%
}-\mathbf{b}(\mathbf{b}\cdot \mbox{curl}\,\mathbf{b})]-\frac{c}{eB}%
\mbox{grad}\,(\mu B+e\phi )\times \mathbf{b}.  \label{dot_X}
\end{equation}%
It follows from Lagrangian (\ref{Lagr_mu}). This formula shows that the
particle moves along the magnetic line (the first term), the second term is
the drift velocity due to the centrifugal force ($\mbox{curl}\,\mathbf{b}-%
\mathbf{b}(\mathbf{b}\cdot \mbox{curl}\,\mathbf{b})=[\mathbf{b}\times (%
\mathbf{b}\cdot \nabla )\mathbf{b}]$ where $(\mathbf{b}\cdot \nabla )\mathbf{%
b}$ is the curvature); the last term is the drift due to the mirror force
and the electric force. Eq. (\ref{dot_X}) can be found in many papers, see,
for example, \cite{Grad1967}.

We consider (quasi-)stationary distributions of the given sort of particles
like that 
\begin{equation}
dN=f\left( \varepsilon (\mathbf{x},\mathbf{v}),\mu (\mathbf{x},\mathbf{v}%
)\right) d\mathcal{V}=[B+\frac{mc}{e}u(\mathbf{b}\cdot \mbox{curl}\,\mathbf{b%
})]F_{\varepsilon }^{\prime }(\varepsilon ,\mu )d^{3}\mathbf{X}dud\mu d\zeta
/(2\pi ),  \label{regular_distr}
\end{equation}%
where $\varepsilon =\mu B+e\phi +mu^{2}/2$ is the Hamiltonian of a guiding
center, and $F(\varepsilon ,\mu )$ is a prescribed function of the two
variables. It is assumed that $F<0$ while $F_{\varepsilon }^{\prime }\propto
f>0$, and $F\rightarrow 0$ as $\varepsilon \rightarrow +\infty $. It is
clear that $f$ satisfies the (collisionless) drift kinetic equation, since
it depends on the exact integral of motion $\varepsilon $ and on the
approximate integral of motion $\mu $ (adiabatic invariant). Therefore there
is no need in checking the hydrodynamic stationarity. We require only two
relations to close the model in a self-consistent manner: they are the
Maxwell equation for a stationary magnetic field and the quasi-neutrality
condition: 
\begin{equation}
\frac{1}{4\pi }\nabla \times \mathbf{B}=\mathbf{j}_{\mathrm{total}}/c,
\label{M_eq}
\end{equation}%
\[
\rho _{\mathrm{total}}=0, 
\]%
where $\mathbf{j}_{\mathrm{total}}$ and $\rho _{\mathrm{total}}$ are the
densities of the electric current and of the electric charge, respectively,
produced by all sorts of particles present in the system.

In the lowest order on gradients, the current density from the given sort of
particles is (it follows from Eq.(3.53) of Ref.\cite{RMP2009}) 
\begin{equation}
\mathbf{j}/c=-\nabla \times (\mathbf{b}N\langle \mu \rangle )+(e/c)N\langle 
\dot{\mathbf{X}}\rangle .  \label{j}
\end{equation}%
Here $N\langle \mu \rangle =|\mathbf{M}|=\int \mu fJdud\mu d\zeta $, where $%
\mathbf{M}$ is the spatial density of the magnetic moment. Using
distribution (\ref{regular_distr}), we have 
\begin{equation}
-\mathbf{b}N\langle \mu \rangle =-\mathbf{B}\int \mu F_{\varepsilon
}^{\prime }(\varepsilon ,\mu )d\mu du=-\mathbf{B}\frac{\partial }{\partial B}%
\int F(\varepsilon ,\mu )d\mu du=\mathbf{B}\frac{\partial }{\partial B}%
\left( \frac{\tilde{p}_{\parallel }}{B}\right) ,  \label{M}
\end{equation}%
where $\tilde{p}_{\parallel }(B,\phi )$ is the parallel pressure of the
given sort of particles, 
\[
\tilde{p}_{\parallel }(B,\phi )=B\int mu^{2}F_{\varepsilon }^{\prime
}(\varepsilon ,\mu )d\mu du=-B\int F(\varepsilon ,\mu )d\mu du. 
\]%
It is remarkable that the calculation of $N\langle \dot{\mathbf{X}}\rangle \equiv
\int \dot{\mathbf{X}}fJdud\mu d\zeta $ with the help of Eqs.(\ref{dot_X})
and (\ref{regular_distr}) results in the following compact expression, 
\begin{equation}
(e/c)N\langle \dot{\mathbf{X}}\rangle \approx \nabla \times (\mathbf{b}%
\tilde{p}_{\parallel }/B).  \label{NU}
\end{equation}

Let us now label each sort of particle by an index $\alpha $. Then the
Maxwell equation (\ref{M_eq}) after substitution of Eqs.(\ref{M}) and (\ref%
{NU}) into Eq.(\ref{j}) for each $\alpha $ and after subsequent summation
over $\alpha $ looks as follows, 
\begin{equation}
\nabla \times \left\{ \mathbf{b}\left[ \frac{B}{4\pi }-\frac{\partial }{%
\partial B}p_{\parallel }(B,\phi )\right] \right\} \approx 0,
\label{equation_B_1}
\end{equation}%
where $p_{\parallel }(B,\phi )=\sum_{\alpha }\tilde{p}_{\alpha }$ is the
total parallel pressure, 
\[
p_{\parallel }(B,\phi )=-\sum_{\alpha }B\int F_{(\alpha )}(\varepsilon
_{\alpha },\mu )d\mu du, 
\]%
with 
\[
\varepsilon _{\alpha }=\mu B+e_{\alpha }\phi +m_{\alpha }u^{2}/2. 
\]%
The quasi-neutrality condition 
\[
\sum_{\alpha }e_{\alpha }B\int \frac{\partial }{\partial \varepsilon
_{\alpha }}F_{(\alpha )}(\varepsilon _{\alpha },\mu )d\mu du=0 
\]%
is easily seen to have the form 
\begin{equation}
\frac{\partial }{\partial \phi }p_{\parallel }(B,\phi )=0.
\label{equation_phi}
\end{equation}%
Since $\mbox{div}\,\mathbf{B}=0$, equations (\ref{equation_B_1}) and (\ref%
{equation_phi}) possess the variational structure, and the corresponding
functional is 
\begin{equation}
\mathcal{F}=\int [B^{2}/(8\pi )-p_{\parallel }(B,\phi )]d^{3}\mathbf{X}.
\label{FreeEnergy}
\end{equation}%
In principle, the quasi-neutrality condition (\ref{equation_phi}) allows one
to express the electric potential $\phi $ through $B$, and then the parallel
pressure in Eq.(\ref{FreeEnergy}) can be understood as a function of $B$
only. As the result, we have the equation 
\begin{equation}
\nabla \times \left\{ \mathbf{b}\left[ \frac{B}{4\pi }-p_{\parallel
}^{\prime }(B)\right] \right\} \approx 0,  \label{equation_B}
\end{equation}%
which coincides with Eq. (\ref{chi}) $\ $at $\chi =0$. Thus, we get a
3D generalization of the 2D variational principle previously derived in \cite%
{KPRS2014} by a different approach.

\bigskip It is worth noting that the quantity 
\[
\frac{4\pi }{c}\mathbf{j=}4\pi \lbrack \nabla \times (\mathbf{b}p_{\parallel
}^{\prime }(B))] 
\]%
can be connected with mean (per volume unit) magnetic moment of plasma $%
\mathbf{M}$=$\mathbf{b}p_{\parallel }^{\prime }(B)$, so that 
the magnetic field $\mathbf{H}=\mathbf{B}+4\pi \mathbf{M}$ \textbf{(}in
accordance with the definition of the Maxwell equations in continuous media. In this case,  equation (\ref{equation_B}) is nothing more as
the Maxwell equation $\nabla \times \mathbf{H=0.}$

Let us consider the most physically interesting case where the functions $%
F_{(\alpha )}(\varepsilon _{\alpha },\mu )$ have the exponential on $%
\varepsilon _{\alpha }$ form, 
\[
F_{(\alpha )}(\varepsilon _{\alpha },\mu )=-\exp (-\varepsilon _{\alpha
}/T_{\alpha })\tilde{D}_{\alpha }(\mu ), 
\]%
with constant temperature parameters $T_{\alpha }$ and some positive
functions $\tilde{D}_{\alpha }(\mu )$. In this case the $u$-integration is
simple, and 
\[
p_{\parallel }(B,\phi )=B\sum_{\alpha }T_{\alpha }\exp (-e_{\alpha }\phi
/T_{\alpha })\int_{0}^{+\infty }\exp (-\mu B/T_{\alpha })D_{\alpha }(\mu
)d\mu , 
\]%
where $D_{\alpha }(\mu )\propto \tilde{D}_{\alpha }(\mu )$ by an $\alpha $%
-dependent factor. Suppose we deal with the simplest electron-proton plasma.
Then 
\[
p_{\parallel }(B,\phi )=T_{i}G_{i}(B)\exp (-e\phi /T_{i})+T_{e}G_{e}(B)\exp
(e\phi /T_{e}), 
\]%
where 
\[
G_{\alpha }(B)=B\int_{0}^{+\infty }\exp (-\mu B/T_{\alpha })D_{\alpha }(\mu
)d\mu ,\qquad \alpha =i,e. 
\]%
The quasi-neutrality condition (\ref{equation_phi}) now takes a simple form, 
\[
\exp (-e\phi /T_{i})G_{i}(B)-\exp (e\phi /T_{e})G_{e}(B)=0, 
\]%
from which we have (compare with Eq. (\ref{total-parallel}) )%
\[
e\phi =\frac{\ln [G_{i}(B)/G_{e}(B)]}{(1/T_{i}+1/T_{e})}, 
\]%
\begin{equation}
p_{\parallel }(B)=(T_{i}+T_{e})[G_{i}(B)]^{\frac{T_{i}}{T_{i}+T_{e}}%
}[G_{e}(B)]^{\frac{T_{e}}{T_{e}+T_{i}}}.  \label{general-parallel}
\end{equation}

In particular, we may assume purely thermal isotropic electron velocity
distribution, which corresponds to $D_{e}(\mu )=const$. In that case $%
G_{e}(B)=const$, and the total parallel pressure simplifies to 
\begin{equation}
p_{\parallel }(B)=n_{0}(T_{i}+T_{e})\left[ \frac{G_{i}(B)}{G_{i}(B_{0})}%
\right] ^{\frac{T_{i}}{T_{i}+T_{e}}}.  \label{P_par_isotr_el}
\end{equation}

\section{Two-dimensional stationary structures of the Grad-Shafranov type}

In two dimensions, we define the stream function $\psi $ (or vector
potential), such that $B_{x}={\partial \psi }/{\partial y}$, $B_{y}=-{%
\partial \psi }/{\partial x}$. In terms of $\psi $ and $B_{z}$, 
\begin{eqnarray}
&&\Big[\Big[\nabla \times \mathbf{B}\Big]\times \mathbf{B}\Big]=\mathbf{e}%
_{x}\Big(-\frac{1}{2}\frac{\partial B_{z}^{2}}{\partial x}-\frac{\partial
\psi }{\partial x}\Delta \psi \Big)  \nonumber \\
&&\qquad +\mathbf{e}_{y}\Big(-\frac{1}{2}\frac{\partial B_{z}^{2}}{\partial y%
}-\frac{\partial \psi }{\partial y}\Delta \psi \Big)-\mathbf{e}_{z}\left\{
\psi ,B_{z}\right\} ,
\end{eqnarray}%
where $\left\{ \psi ,B_{z}\right\} $ denotes the Jacobian. Furthermore, $%
\nabla _{\perp }=\nabla -\frac{1}{B^{2}}\mathbf{B}_{\perp }\mathbf{(B_{\perp
}\cdot \nabla )-}\frac{B_{z}}{B^{2}}\mathbf{e}_{z}(\mathbf{B}_{\perp }\cdot
\nabla )$, where $\nabla \equiv (\partial _{x},\partial _{y})$ and $\mathbf{B%
}_{\perp }=(B_{x},B_{y})$.

In Eq. (\ref{MHD2}), we now separate the $(x,y)$-components: 
\begin{eqnarray}
&&-\nabla p_{\perp }+\frac{1}{B^{2}}\mathbf{B}_{\perp }\mathbf{(\mathbf{B}%
_{\perp }\cdot \nabla )}p_{\perp }  \nonumber  \label{perp1} \\
&&\quad +\frac{1}{2B^{2}}(p_{\perp }-p_{\Vert })\left[ \nabla -\frac{1}{B^{2}%
}\mathbf{B}_{\perp }\mathbf{(\mathbf{B}_{\perp }\cdot \nabla )}\right] B^{2}
\\
&&\quad +\frac{1}{4\pi }\left[ 1+\frac{4\pi }{B^{2}}(p_{\perp }-p_{\Vert })%
\right] \left( -\frac{1}{2}\nabla B_{z}^{2}-\nabla \psi \Delta \psi \right)
=0.  \nonumber
\end{eqnarray}%
Due to identity (\ref{identity3}), the equation for the $z$ component can be
written 
\begin{eqnarray}
&&\frac{B_{z}}{4\pi }\left[ (\mathbf{B}_{\perp }\cdot \nabla )\left( 1+\frac{%
4\pi }{B^{2}}(p_{\perp }-p_{\Vert })\right) \right]  \nonumber \\
&&+\frac{1}{4\pi }\left[ 1+\frac{4\pi }{B^{2}}(p_{\perp }-p_{\Vert })\right]
(\mathbf{B}_{\perp }\cdot \nabla )B_{z}=0.
\end{eqnarray}%
In terms of $\psi $, after integration, it leads to 
\begin{equation}
\frac{B_{z}}{4\pi }\left( 1+\frac{4\pi }{B^{2}}(p_{\perp }-p_{\Vert
})\right) =f(\psi ).  \label{B-z1}
\end{equation}%
Interestingly, in the isotropic case ($p_{\perp }-p_{\Vert }=0)$, we have $%
B_{z}=B_{z}(\psi ),$ in full agreement with the Grad-Shafranov reduction 
\cite{grad,shafranov-58,shafranov-66}. Furthermore, because the projection
of the full equation on $\mathbf{B}$ is equal to zero, in the 2D case where
the fields are functions of $x$ and $y$ only, the projection of Eq. (\ref%
{perp1}) on $\mathbf{B_{\perp }}$ vanishes identically. Therefore the
relevant information is obtained by taking the vector product of Eq. (\ref%
{perp1}) with $\mathbf{B_{\perp }}$, in the form 
\begin{eqnarray}
&&\left( \nabla \psi \cdot \nabla \left[ p_{\perp }+\frac{B_{z}^{2}}{8\pi }%
\right] \right) -\frac{(p_{\perp }-p_{\Vert })}{2B^{2}}\left( \nabla \psi
\cdot \nabla \left( B^{2}-B_{z}^{2}\right) \right)  \nonumber \\
&&\qquad =-\frac{\left( B^{2}-B_{z}^{2}\right) }{4\pi }\left[ 1+\frac{4\pi }{%
B^{2}}(p_{\perp }-p_{\Vert })\right] \Delta \psi .  \label{analog}
\end{eqnarray}%
This equation is supplemented by relation (\ref{B-z1}).

Equation (\ref{analog}) can be viewed as analogous to the Grad-Shafranov
equation, the main difference being that the pressures are here prescribed
as functions of the magnetic field amplitude. In particular, it does not
reduce in the isotropic case to the usual Grad-Shafranov equation. Note that,
according to the previous section, the obtained equations (\ref{B-z1}) and (%
\ref{analog}) follow from the variational principle \ for $\mathcal{F}$. 
In particular, for the purely two-dimensional geometry when $B_{z}=0$ and $%
B^{2}=|\mathbf{B}_{\perp }|^{2}$   Eq. (\ref{analog}) reduces to 
\begin{equation}
\nabla \cdot \left\{ \left[ 1+\frac{4\pi }{B^{2}}(p_{\perp }-p_{\Vert })%
\right] \nabla \psi \right\} =0  \label{Eq-psi}
\end{equation}%
and thus derives from the variational principle $\delta \mathcal{F}=0$ with $%
\mathcal{F}=\frac{1}{4\pi }\int g(|\nabla \psi |^{2})dxdy$. Here the
function $g$ is found by integrating 
\[
g^{\prime }(B^{2})=1+\frac{4\pi }{B^{2}}(p_{\perp }-p_{\Vert }).
\]%
Due to identity (\ref{identity3}), we have 
\begin{equation}
\mathcal{F}=\int \left( \frac{B^{2}}{8\pi }-p_{\Vert }\right) \,dx\,dy\equiv
-\int \Pi _{\Vert }\,dx\,dy.  \label{free-general}
\end{equation}%
It follows that all the two-dimensional stationary states in anisotropic MHD
are stationary points of the functional $\mathcal{F}$. Its density is a
function of $B=|\nabla \psi |$ only. In the special case of cold electrons,
this free energy turns out to identify with the Hamiltonian of the static
problem \cite{PRS06}.

Equations similar to (\ref{Eq-psi}) arise in the context of pattern
structures in thermal convection. As shown in \cite%
{ErcolaniIndikNewellPassot}, such equations represent integrable
hydrodynamic systems. As in the usual one-dimensional gas dynamics, these
systems display breaking phenomena where the solution looses its smoothness
at finite distance, due to the formation of folds. As a consequence, these
models require some regularization. For patterns, the authors of \cite%
{ErcolaniIndikNewellPassot} supplement in the equation an additional linear
term involving a square Laplacian. In our case, this procedure corresponds
to the replacement of $\mathcal{F}$ by $\mathcal{F}+({\nu }/{2})\int \left(
\Delta \psi \right) ^{2}dxdy$, with a constant $\nu >0$. In plasma physics,
regularization can originate from finite Larmor radius (FLR) corrections,
which are not retained in the present analysis based on the drift kinetic
equation (see, e.g. \cite{KPS2007a, KPS2007b}). \ In the three-dimensional
geometry the same regularization reads as (compare to \cite{KPRS2014}): 
\begin{equation}
\tilde{\mathcal{F}}=\int \left[ \frac{B^{2}}{8\pi }-p_{\parallel }(B)+\frac{%
\nu }{2}|\nabla \times {\mathbf{B}}|^{2}\right] d^{3}\mathbf{r}.
\label{FreeEnergy_reg}
\end{equation}%
One more remark. Let $\mathbf{B}$ be a function of $x$ and $y$, but $%
B_{z}\neq 0$. In this case $B_{z}$ is not defined by stream function $\psi $ and therefore  one needs to write down 
\begin{equation}
\left[ \mathbf{B}\times \left[ \nabla \times \frac{\delta \mathcal{F}}{%
\delta \mathbf{B}}\right] \right] =0.  \label{var2}
\end{equation}%
Hence it is easily to get Eqs. (9,10) from \cite{KPRS2014}. It is necessary
to mention also that for the 2D case the stationary states with $B_{z}\neq 0$
are determined from the equation%
\[
\left[ \nabla \times \frac{\delta \mathcal{F}}{\delta \mathbf{B}}\right] =0.
\]%
For instance, the equation for $B_{z}$ has the form 
\[
\left( 1+4\pi \frac{p_{\perp }-p_{\parallel }}{B^{2}}\right) B_{z}=%
\mbox{const},
\]%
where instead of arbitrary function of $\psi $ (see Eq. (9) from \cite%
{KPRS2014}) we have const. In $r-z$ geometry the analogous situation takes
place where $B_{\varphi }$ plays the same role as $B_{z}$ in the planar case.

Note that for isotropic plasma ($p_{\perp }=p_{\Vert }$), H. Grad and H.
Rubin \cite{GradRubin1958} formulated for the stationary MHD states the
variational principle for 
\[
\mathcal{F}=\int \left( \frac{B^{2}}{8\pi }-p\right) d\mathbf{r}.
\]

\subsection{KP soliton}

We shall  now show that the functional $\mathcal{F}$ we previously
introduced has the meaning of a free energy. In the weakly nonlinear regime
near the MI threshold, the temporal behavior of the mirror modes can be
described by a 3D model \cite{KPS2007a, KPS2007b, KPS2012a}, that in the
present 2D geometry reads 
\begin{equation}
u_{t}=-\widehat{|k_{y}|}\frac{\delta F}{\delta u}  \label{asymp}
\end{equation}%
with the free energy 
\begin{equation}
F=\int \left[ \frac{1}{2}(-\varepsilon u^{2}+u\frac{\partial _{z}^{2}}{%
\Delta _{\perp }}u+\left( \nabla _{\perp }u\right) ^{2})+\frac{\lambda }{3}%
u^{3}\right] d\mathbf{r.}  \label{F-3D}
\end{equation}%
Here $u$ denotes the dimensionless magnetic field fluctuations and $%
\varepsilon $ the distance from MI threshold. The third term in $F$
originates from the FLR corrections, and $\lambda $ is a nonlinear coupling
coefficient which is positive for bi-Maxwellian distributions. In Eq. (\ref%
{asymp}), the operator $\widehat{|k_{y}|}$ is a positive definite operator
(in the Fourier representation it reduces to $|k_{y}|$), so that Eq. (\ref%
{asymp}) has a generalized gradient form.

Let us now show that this result can be obtained from the functional $%
\mathcal{F}$ defined in (\ref{free-general}). We isolate the perturbation $%
\varphi $ in the stream function $\psi =-B_{0}(x+\varphi )$ with $\varphi
\rightarrow 0$ as $|\mathbf{r}|\rightarrow \infty $, so that the mean
magnetic field $\mathbf{B}_{0}$ is directed along the $y$-axis. We then
expand Eq. (\ref{free-general}) in series with respect to $u$. For the sake
of simplicity, we restrict the analysis to the case of cold electrons. The
expansion of the integrand $B^{2}/(8\pi) -p_{\Vert }$ in $F$ has then the
form 
\begin{eqnarray}
&&n_{0}T_{\parallel }\left[ \frac{\left( u+1\right) ^{2}}{\beta _{\Vert }}-%
\frac{1+u}{1+au}\right]  \nonumber \\
&&\qquad =n_{0}T_{\parallel }[ \left( \beta _{\Vert }^{-1}-1\right) +u\left(
a+2\beta _{\Vert }^{-1}-1\right)  \nonumber \\
&&\qquad +u^{2}\left( -a^{2}+a+\beta _{\Vert }^{-1}\right) -u^{3}
a^{2}\left( a-1\right) +....]  \label{expansion}
\end{eqnarray}
where we use the usual notation $\beta _{\Vert }=8\pi
n_{0}T_{\parallel}/B_{0}^{2}$.

As well known (see, e.g. \cite{KPS2007a, KPS2007b}), near threshold, MI
develops in quasi-transverse directions relative to $\mathbf{B}_{0}$. This
means that, in the 2D geometry, $\varphi _{x}\gg \varphi _{y}$ and, with a
good accuracy, $u$ coincides with $\varphi _{x}$. However, in the expansion
of $u=\sqrt{(\varphi _{x}+1)^{2}+\varphi _{y}^{2}}-1 \simeq \varphi _{x}+
\varphi_{y}^{2}/{2}$, it is necessary to keep the second term, quadratic
with respect to $\varphi $. The linear term in the expansion of $\mathcal{F}$
vanishes and the quadratic terms is given by 
\[
\mathcal{F}_{2}=n_{0}T_{\parallel }\int \Big\{ \Big[ a(a-1)+\frac{1}{\beta}
_{\Vert }\Big] \varphi _{x}^{2}+\Big[ a-1+\frac{2}{\beta} _{\Vert }\Big ]
\varphi _{y}^{2}\Big \} dxdy. 
\]%
where the factor $a(a-1)+1/\beta \equiv -\varepsilon /2$ defines the MI
threshold $a=1+{1}/{\beta _{\perp }}$ (that the present equations of state
accurately reproduces). It is also seen that for $|\varepsilon| \ll 1$, $%
\varphi _{x}/\varphi _{y}\sim |\varepsilon |^{-1/2}$, in agreement with the
quasi-one-dimensional development of MI near threshold. In this case, $%
\mathcal{F}_{2}$ coincides with the quadratic term in (\ref{F-3D}), up to a
simple rescaling and to the FLR contribution, Furthermore, the cubic term in
(\ref{expansion}) gives the nonlinear coupling coefficient $\lambda =
a\left( a-1\right) >0$. As a consequence, $\mathcal{F}$, introduced in the
previous section, reduces to the free energy of the asymptotic model. The
temporal equation for $\varphi $ has also the generalized gradient form
originating from (\ref{asymp}), 
\begin{equation}
\varphi _{t}=-\Gamma \frac{\delta F}{\delta \varphi} \ \text{with} \ \Gamma
= -\widehat {\frac{|k_{y}|}{k_{x}^{2}}},  \label{phi-F}
\end{equation}
for which the associated stationary equation reads 
\begin{equation}
\varepsilon \varphi _{xx}+\varphi _{xxxx}-\varphi _{yy}-\lambda \partial
_{x}\left( \varphi _{x}^{2}\right) =0,  \label{stat-phi}
\end{equation}%
where the linear operator $L=-\varepsilon \partial _{xx}+\partial
_{yy}-\partial _{xxxx}$ is elliptic or hyperbolic depending on the sign of $%
\varepsilon $. For $\varepsilon >0$ (above threshold), this operator is
hyperbolic, while below threshold it is elliptic and thus invertible in the
class of functions vanishing at infinity. Remarkably, in the latter case,
Eq. (\ref{stat-phi}) identifies with the soliton for KP equation called
lump. In standard notations, lump is indeed a solution of the stationary
KP-II equation, 
\begin{equation}
-Vu_{xx}+u_{xxxx}-u_{yy}+3(u^{2})_{xx}=0,  \label{KP}
\end{equation}%
where $V$ is the lump velocity. When comparing this equation with (\ref%
{stat-phi}) we see that $-|\varepsilon |$ plays the role of the lump
velocity $V $ and $\lambda \varphi _{x}\rightarrow -3u$.

The lump solution was first discovered numerically by Petviashvili \cite%
{Petviashvili} using the method now known as the Petviashvili scheme (see
the next section). The analytical solution was later on obtained in \cite%
{BIMMZ}. In our notation, it reads%
\[
\varphi _{x}=-\frac{12|\varepsilon |}{\lambda }\frac{\left( 3+ \varepsilon
^{2}y^{2}-|\varepsilon |x^{2}\right) }{\left[ 3+ \varepsilon
^{2}y^{2}+|\varepsilon |x^{2}\right] ^{2}}. 
\]%
This function vanishes algebraically at infinity like $r^{-2}$. In the
center region $-|\varepsilon |^{-2}\sqrt{|\varepsilon |x^{2}-3}%
<y<|\varepsilon |^{-2}\sqrt{|\varepsilon |x^{2}-3}$, the magnetic field
displays a hole with a minimum at $x=y=0$ equal to $-4|\varepsilon |/\lambda$%
. In the outer region, the magnetic lump has two symmetric humps with
maximum values $|\varepsilon|/(2\lambda)$ at $y=0$ and $x=\pm 3|\varepsilon
|^{-1/2}$. The main contribution to the ``skewness'' $I=\int \varphi _{x}
^{3} \,dx\,dy$ comes from the hole region, providing a negative value to $I$%
, in complete agreement with \cite{KPS2007a, KPS2007b}.

\noindent 

\section{Numerical 2D solutions}

In the 2D case, our regularized model equation for stationary
pressure-balanced structures has a variational form 
\begin{equation}
-\partial _{x}\left[ \frac{(1+\varphi _{x})}{(1+u)}\frac{dg}{du}\right]
-\partial _{y}\left[ \frac{\varphi _{y}}{(1+u)}\frac{dg}{du}\right] +\nu
\Delta ^{2}\varphi =0.  \label{variational_form_reg}
\end{equation}%
Clearly, Eq. (\ref{variational_form_reg}) describes stationary points $%
\delta \mathcal{F}/\delta \varphi =0$ of the functional $\mathcal{F}=\int
[g(u)+(\nu /2)(\Delta \varphi )^{2}]\,dx\,dy$, with some constant parameter $%
\nu $ (in this expression and everywhere below, we use dimensionless
variables).

We applied two numerical methods to solve Eq. (\ref{variational_form_reg}).
The first one is a generalization of the well known gradient method which
corresponds to a dissipative dynamics along an auxiliary time-like variable $%
\tau $ of the form $\varphi _{\tau }=-\widehat{\Gamma} ({\delta {\mathcal{F}}%
}/{\delta \varphi})$, with a positive definite linear operator $\widehat{%
\Gamma}$. It is clear that attractors in the phase space of the above
dynamical system are stable solutions of Eq. (\ref{variational_form_reg}).
Unstable solutions however cannot be found by this method.

Furthermore, the linear part of Eq. (\ref{variational_form_reg}) is of the
form ${\widehat L} \varphi= -g^{\prime \prime }(0) \varphi _{xx}-g^{\prime
}(0)\varphi _{yy}+\nu \Delta^2 \varphi$. The coefficient $g^{\prime \prime
}(0)$ is proportional to $\varepsilon$ (introduced in the previous section)
and $g^{\prime }(0)$ is positive within the adiabatic approximation. When
these two are positive, the operator $L$ is elliptic and it is possible to
employ the so-called Petviashvili method \cite{Petviashvili}. It is a
specific method for finding localized solutions of equations of the form $%
\widehat{M}\varphi =N[\varphi], $ with a positively definite linear operator 
$\widehat{M}$ and a nonlinear part $N[\varphi ]$. Note that in our case the
Fourier image of $\widehat{M}$ is 
\begin{equation}
M(k_{x},k_{y})= g^{\prime \prime }(0)k_{x}^{2}+g^{\prime }(0)k_{y}^{2}+\nu
(k_{x}^{2}+k_{y}^{2})^{2}>0.
\end{equation}%
In its simplest form, the iteration scheme of the Petviashvili method reads 
\begin{equation}
\varphi _{n+1}=(\widehat{M}^{-1}N[\varphi _{n}])\left( \frac{\int \varphi
_{n}\widehat{M}\varphi _{n}\,dx\,dy}{\int \varphi _{n}N[\varphi _{n}]\,dx\,dy%
}\right) ^{-\gamma },
\end{equation}%
where $\gamma $ is a positive parameter in the range $1<\gamma <2$. The
corresponding multiplier strongly affects the structure of attractive
regions in the phase space.

It is worth noting that if the operator ${\widehat L}$ is hyperbolic,
solutions of the problem are not localized with respect to both $x$ and $y$
coordinates, and will be periodic or more generally quasiperiodic \cite{ZK,
KD}.

\begin{figure}[tbp]
\begin{center}
\includegraphics[width=0.45\textwidth]{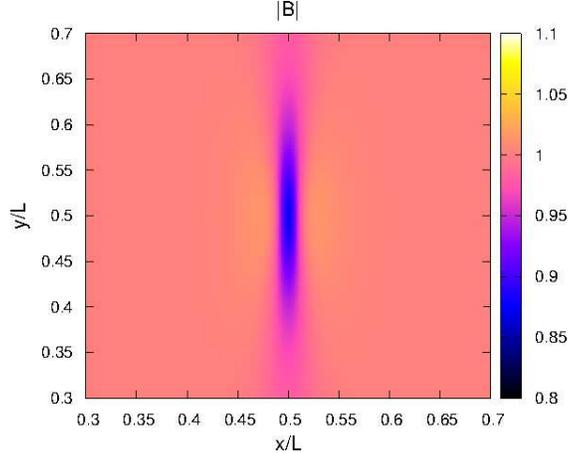}
\end{center}
\caption{Fig. 1. Unstable localized solution for $\protect\nu=0.0004$, $r=7$%
, $B_{s}=0.5$ (in units of $B_{0}$), and $C=0.002$. The value $1/\protect%
\beta_\parallel=1.127$ prescribes an aspect ratio $\protect\sqrt{g^{\prime
\prime }(0)/g^{\prime }(0)}=0.2$.}
\label{lump}
\end{figure}

\begin{figure}[tbp]
\begin{center}
\includegraphics[width=0.4\textwidth]{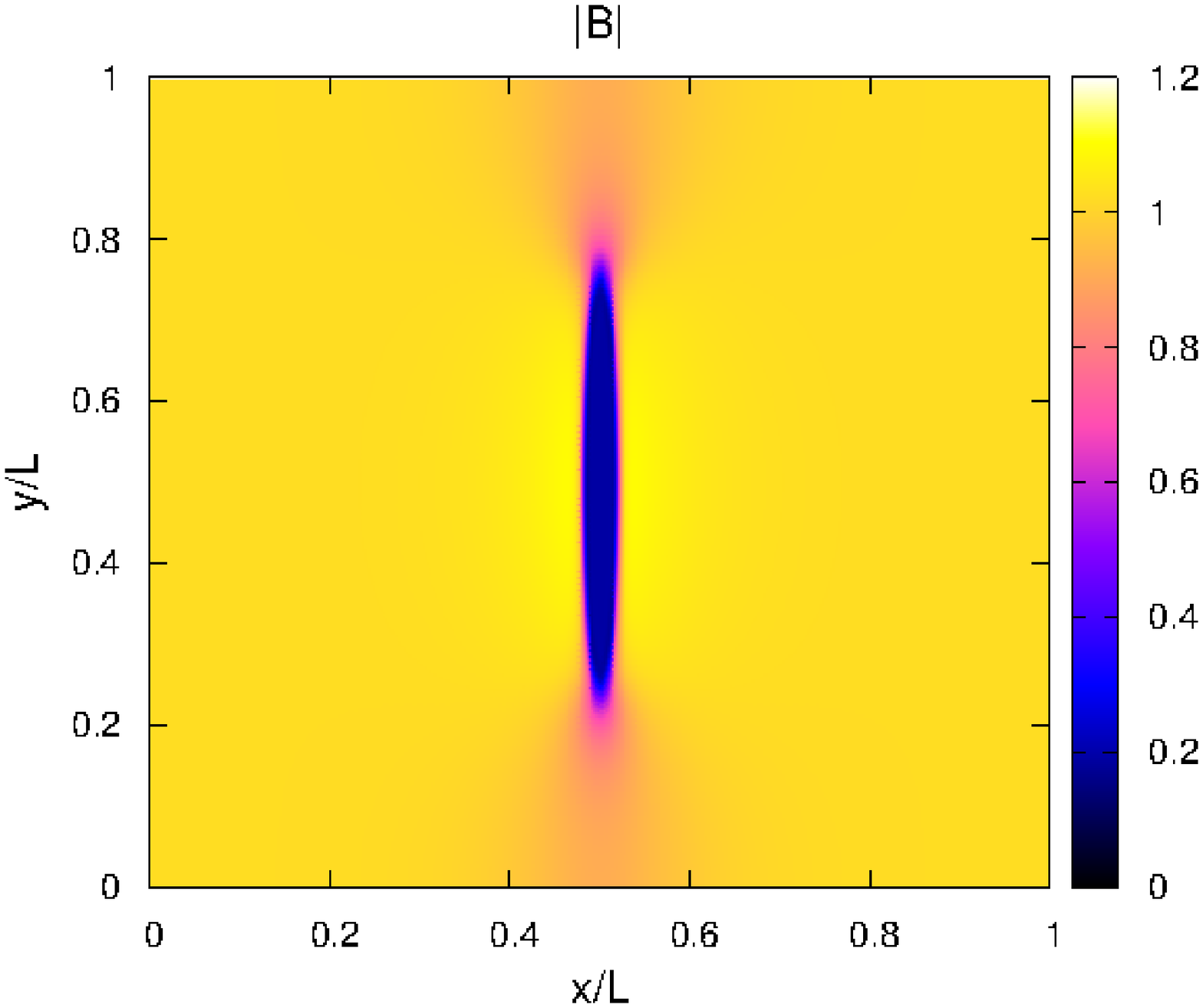}\\[0pt]
\vspace{4mm} \includegraphics[width=0.4\textwidth]{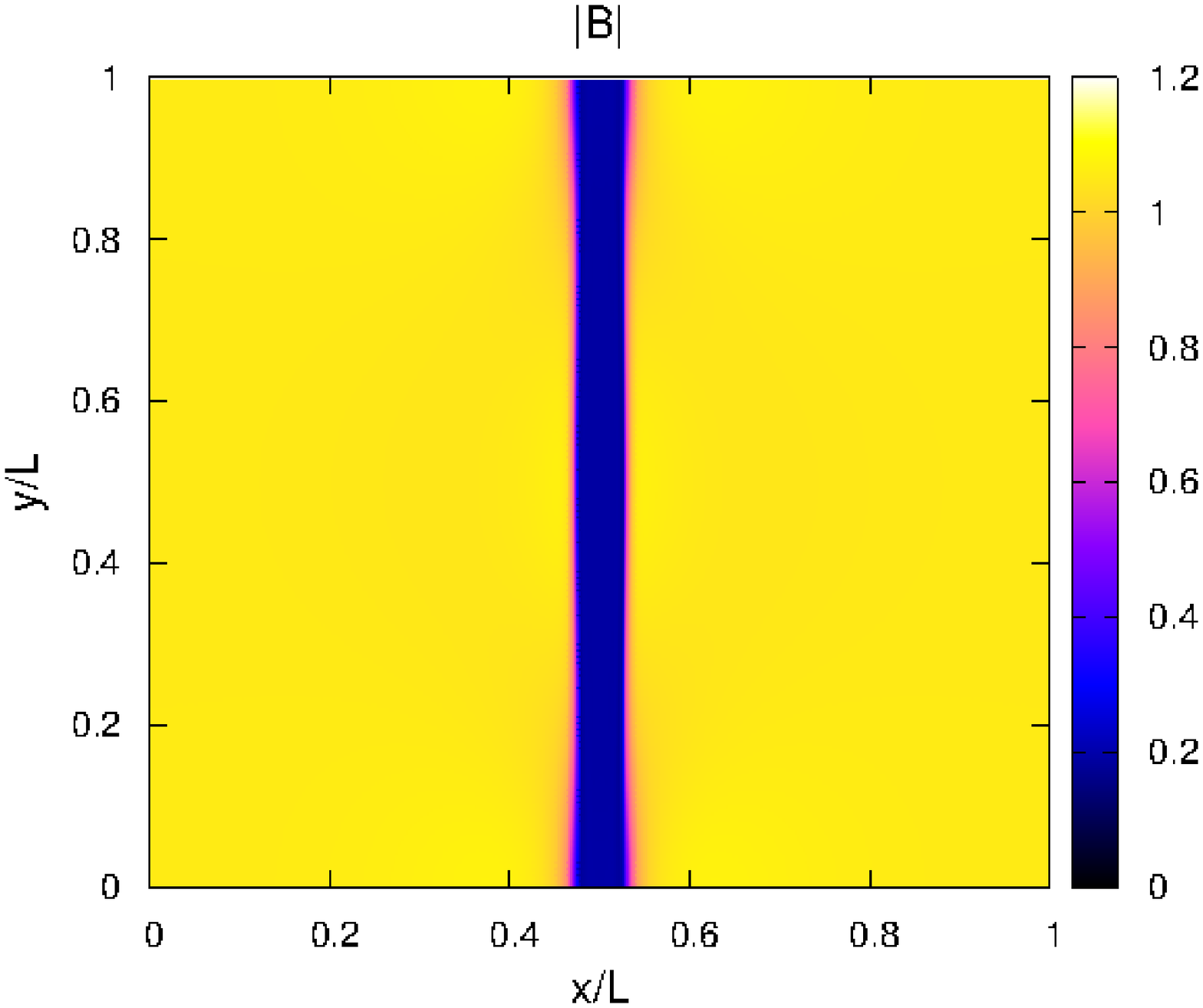}\\[0pt]
\vspace{4mm} \includegraphics[width=0.4\textwidth]{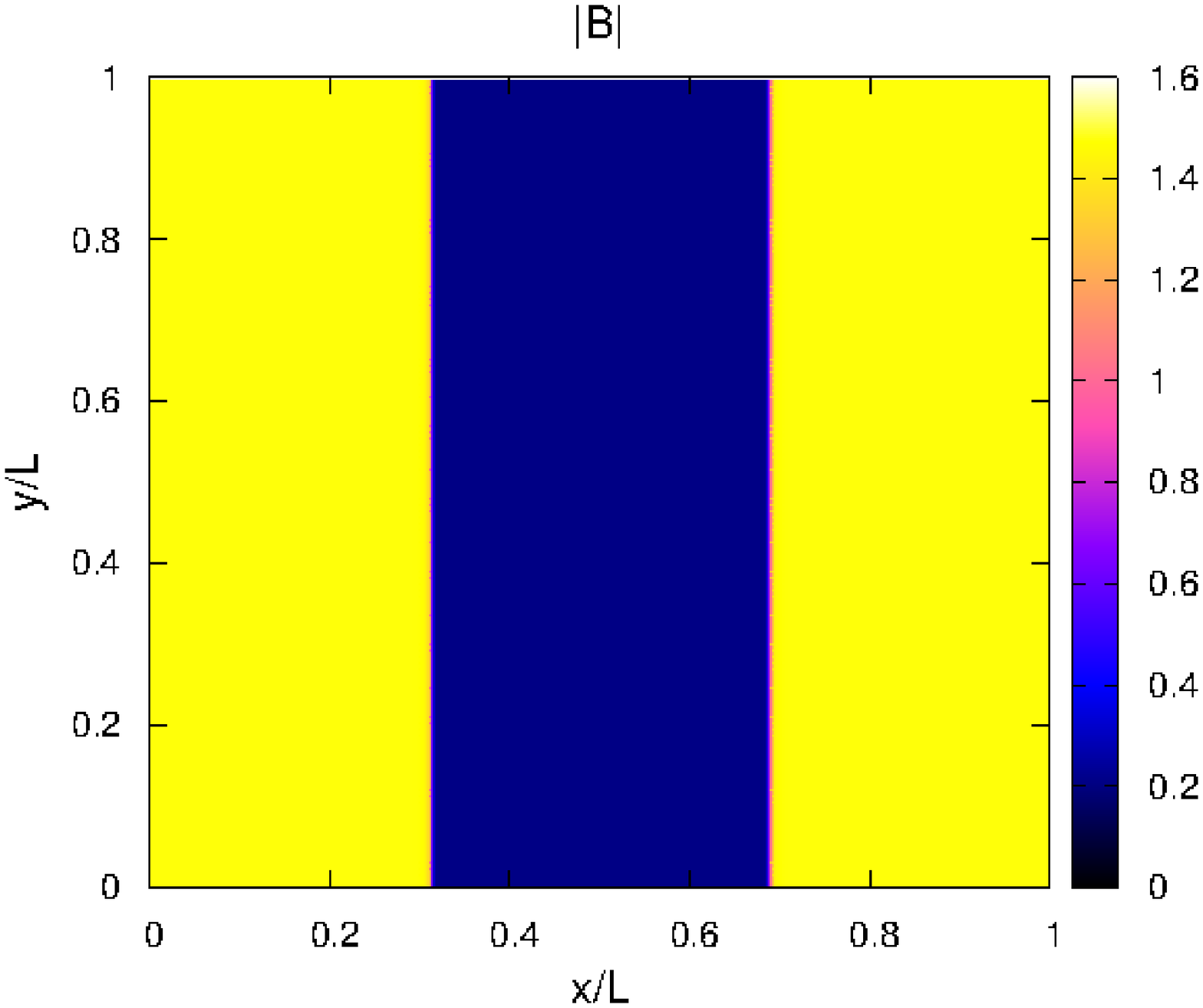}
\end{center}
\caption{Fig. 2. Formation of a stable 1D solution in a gradient
computation, for the same parameters as in Fig.1.}
\label{stripe}
\end{figure}

\subsection{The results}

We performed computations with both numerical methods using fast Fourier
transform numerical routines for the evaluation of the linear operators.
Periodic boundary conditions for a computational square $2\pi \times 2\pi $
were assumed.

For the gradient method, we used the simplest first-order Euler scheme for
stepping along $\tau$, with $\delta\tau\sim 0.01 $. The operator $\widehat
\Gamma$ was taken in a form giving stable computation, namely $%
\Gamma(k_x,k_y)=1/[k_x^2+k_y^2+ \nu(k_x^2+k_y^2)^2]$.

As for the Petviashvili method, the value $\gamma=1.8$ was used, leading,
after an erratic transient, to a convergence of the iterations to unstable
solutions of the variational equation (\ref{variational_form_reg}).

The main results of our computations can be formulated as follows. There do
exist unstable localized solutions of Eq. (\ref{variational_form_reg}),
which are similar to the lump solutions of KPII equation, when written in
terms of $u=\partial_x\varphi$ (Fig. \ref{lump}). For asymptotically small $%
\varepsilon$, they accurately coincide with KP solutions, independently of
the electron temperature, as it should be. Such low-amplitude stationary
states do not depend on the particular choice of the regularization of $g(u)$%
. No other kinds of solutions were found with the Petviashvili method.

When the gradient method is used, large amplitudes $u\sim 1$ are achieved in
many cases, and the final result turns out to be dependent on the choice of
the parameters $r$ and $C$ in the regularized function $g$. Without
regularization, no smooth stationary state is approached. Instead, a
singularity occurs. Differently, when a regularized $g$ with parameters $%
r\sim 10$ and $C\sim 0.001$ is used, the final state identifies with a
one-dimensional stripe in the form of a magnetic hole, as shown in Fig. \ref%
{stripe} that also displays typical stages of the ``gradient'' evolution. In
all simulations, the magnetic field in the stripe was smaller than the
`singular' magnetic field $B_{s}$ given by Eq. (\ref{singularB}). For
increasing $r$, the magnetic field in the stripe tends to decrease, down to
0. For initial conditions in the form of a slightly perturbed 2D lump, the
final result is always a one-dimensional stripe of hole type, which
demonstrates the instability of the 2D lump, in full agreement with the
analytical prediction \cite{KPS2007a, KPS2007b}.

In no cases stable 2D structures localized both in $x$ and $y$ directions
were found. Instead, the gradient method showed that stable structures can
only be one-dimensional, transverse to the magnetic field. An initial
localized perturbation of sufficiently high amplitude develops into an
increasingly long structure along the $y$ axis, and eventually reaches the
boundary of the computational domain.

The question arises whether the 1D shock solutions obtained in \cite{PRS06}
(for which $\mathrm{min}B>B_{s}$) would identify with the present solution
when $\nu \rightarrow 0$, a limit which is unreachable in the present
numerics. It is possible that the presence of the bi-Laplacian
regularization leads to overshooting in the shock solution, resulting in the
convergence towards solutions where $\mathrm{min}B<B_{s}$.

\subsection{2D mirror structures with $B_{z}\not=0$: stripes and magnetic
bubbles}

\begin{figure}[tbp]
\begin{center}
\epsfig{file=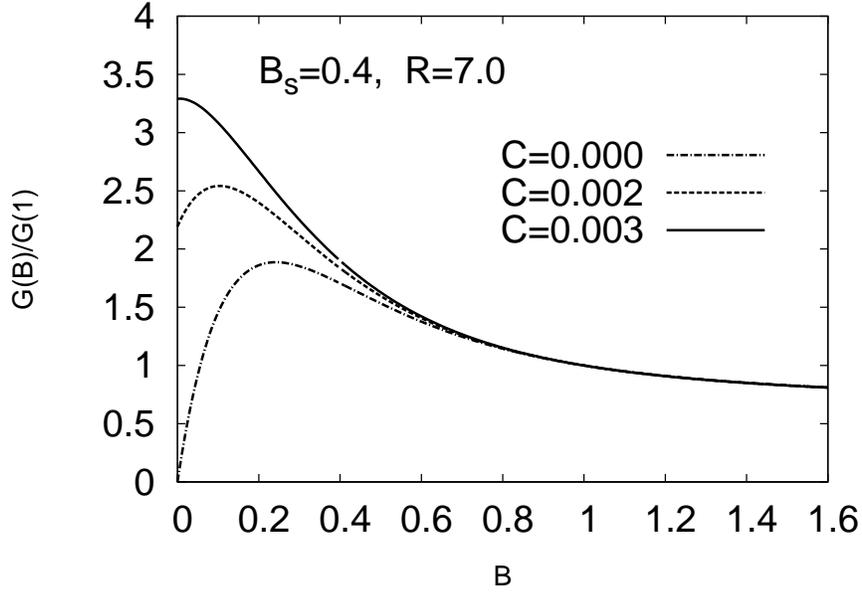,width=120mm}
\end{center}
\caption{Some plots corresponding to expression (\protect\ref{W_reg}).}
\label{W_plot}
\end{figure}

Let us consider some numerical examples. For simplicity we take the function 
$D_i(\mu) \propto \exp(\mu B_s/T_i)$ for $\mu<\mu_*$, and $D_i(\mu)=const$
for $\mu>\mu_*$, with some parameter $B_s<B_0$, and a large $\mu_*$. Such
constant-like behaviour of $D_i(\mu)$ at very large $\mu$ is necessary both
from formal and physical points of view (see discussion in Ref.\cite%
{KPRS2014}). At $B=B_0$ we thus have a nearly Gaussian ion perpendicular
velocity distribution with the temperature $T_\perp(B_0)=T_i/(1-B_s/B_0)>T_i$%
. The distribution becomes strongly non-Gaussian as the magnetic field
decreases to values $B\lesssim B_s$. Let us normalize all magnetic field
values to $B_0$ so that formally $B_0=1$. As the result, we have the
following expression for the ratio $G_i(B)/G_i(1)\equiv W(B)$, 
\begin{equation}
W(B) =\frac{B(1-C)(1-B_s)}{\{1-\exp[-R(1-B_s)]\}} \frac{\{1-\exp[-R(B-B_s)]\}%
}{(B-B_s)} +C\exp[R(1-B)],  \label{W_reg}
\end{equation}
with a sufficiently large regularizing parameter $R$ and a small parameter $%
C $. Some plots, with $B_s=0.4$, $R=7.0$, for several $C$, are shown in Fig.%
\ref{W_plot} 

\begin{figure}[tbp]
\begin{center}
\epsfig{file=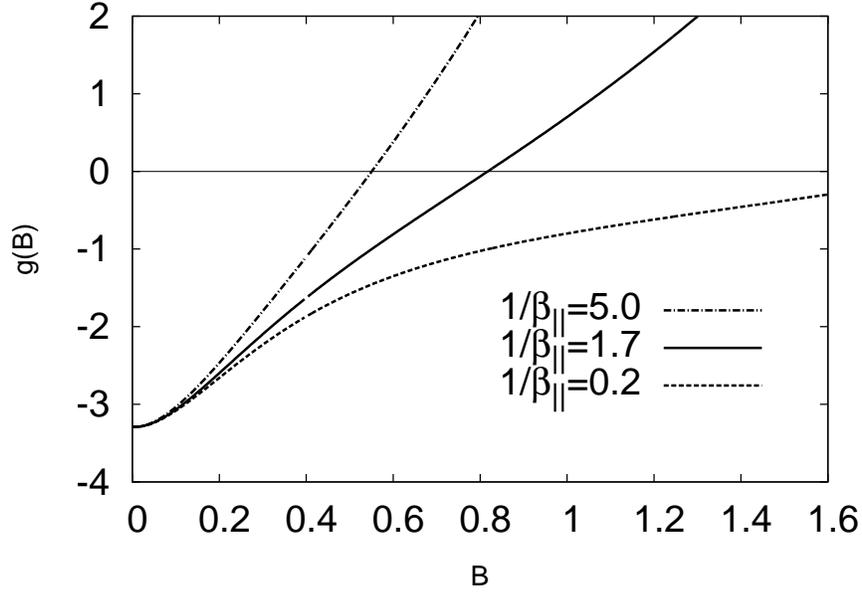,width=120mm}
\end{center}
\caption{Some examples corresponding to expression (\protect\ref{g}). Here $%
B_{s}=0.4$, $R=7.0$, $C=0.003$. System with $1/\protect\beta _{\parallel
}=0.2$ is linearly unstable. System with $1/\protect\beta _{\parallel }=1.7$
is linearly stable, but subcritical structures are possible. System with $1/%
\protect\beta _{\parallel }=5.0$ is stable, no structures are possible.}
\label{g_plot}
\end{figure}

We substituted this dependence into Eq.(\ref{P_par_isotr_el}) and then into
Eq.(\ref{FreeEnergy_reg}), with $T_{e}=0$. To find stable stationary 2D
mirror structures with $B_{z}\not=0$, we parametrized magnetic field in the
following manner, 
\[
B_{x}=-\psi _{y}(x,y),\qquad B_{y}=\psi _{x}(x,y),\qquad B_{z}=\gamma (x,y). 
\]%
We fixed mean values $\langle B_{x}\rangle =0$, $\langle B_{y}\rangle =\cos
\Theta _{0}$, $\langle B_{z}\rangle =\sin \Theta _{0}$. Then we employed the
gradient numerical method described in Ref.\cite{KPRS2014} [with a simple
generalization to include $\gamma (x,y)$] to find minimum of the functional 
\begin{equation}
\tilde{\mathcal{F}}_{2D}=\int \Big[g\Big(\sqrt{|\nabla \psi |^{2}+\gamma ^{2}%
}\Big)+\frac{\nu }{2}(|\Delta \psi |^{2}+|\nabla \gamma |^{2})\Big]d^{2}%
\mathbf{X},  \label{F_psi_gamma}
\end{equation}%
where 
\begin{equation}
g(B)=\frac{B^{2}}{\beta _{\parallel }}-W(B),  \label{g}
\end{equation}%
and $\beta _{\parallel }=8\pi n_{0}T_{i}/B_{0}^{2}$. Plots of function $g(B)$%
, for several values of $\beta _{\parallel }$, are shown in Fig.\ref{g_plot}%
. The mirror instability takes place when the second derivative $g^{^{\prime
\prime }}(1)$ is negative. Subcritical mirror structures are possible when $%
g^{^{\prime \prime }}(1)$ is positive, but there is a range of $B$ where $%
g^{^{\prime \prime }}(B)<0$. 

\begin{figure}[tbp]
\begin{center}
\epsfig{file=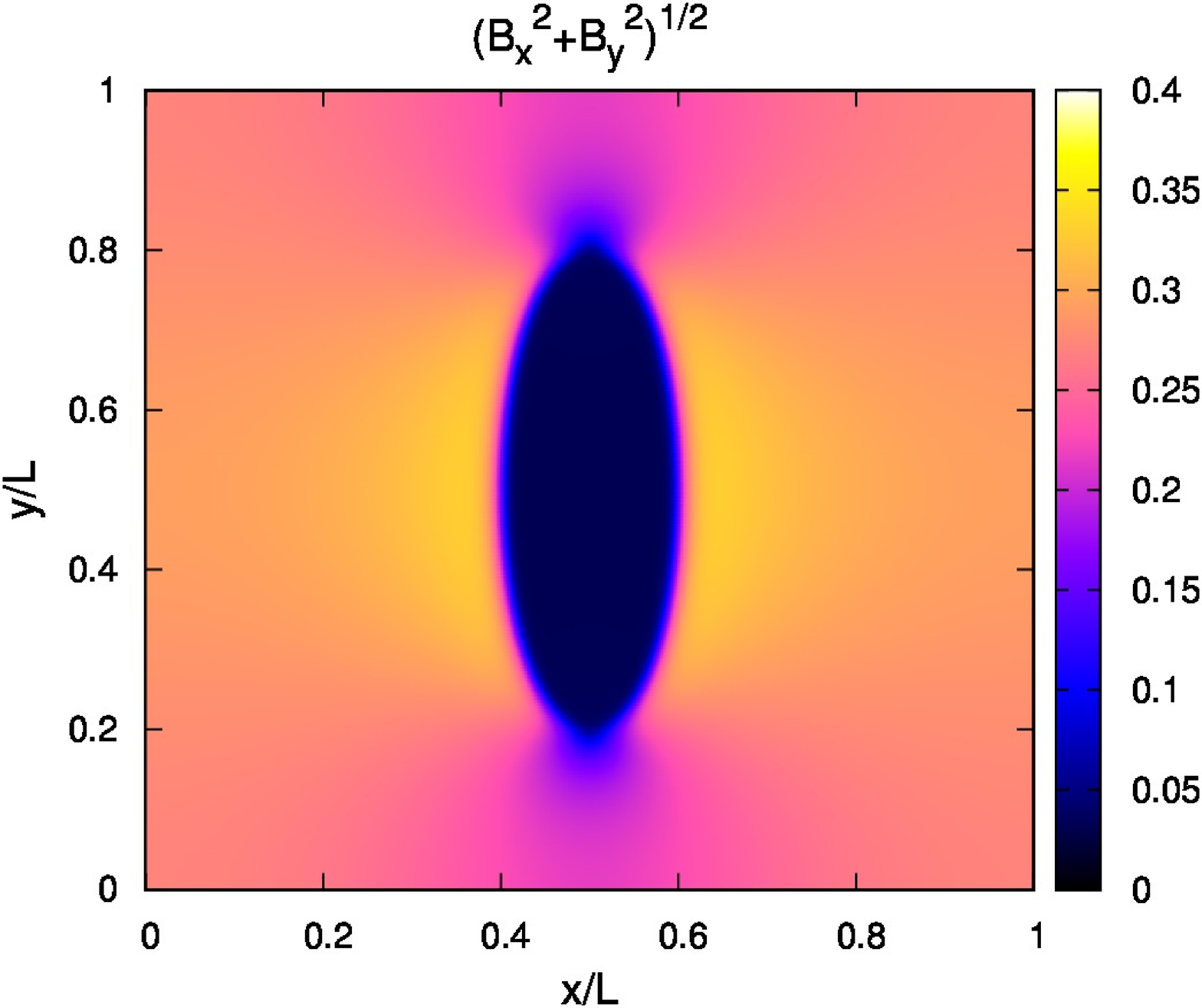,width=110mm}\\[0pt]
\epsfig{file=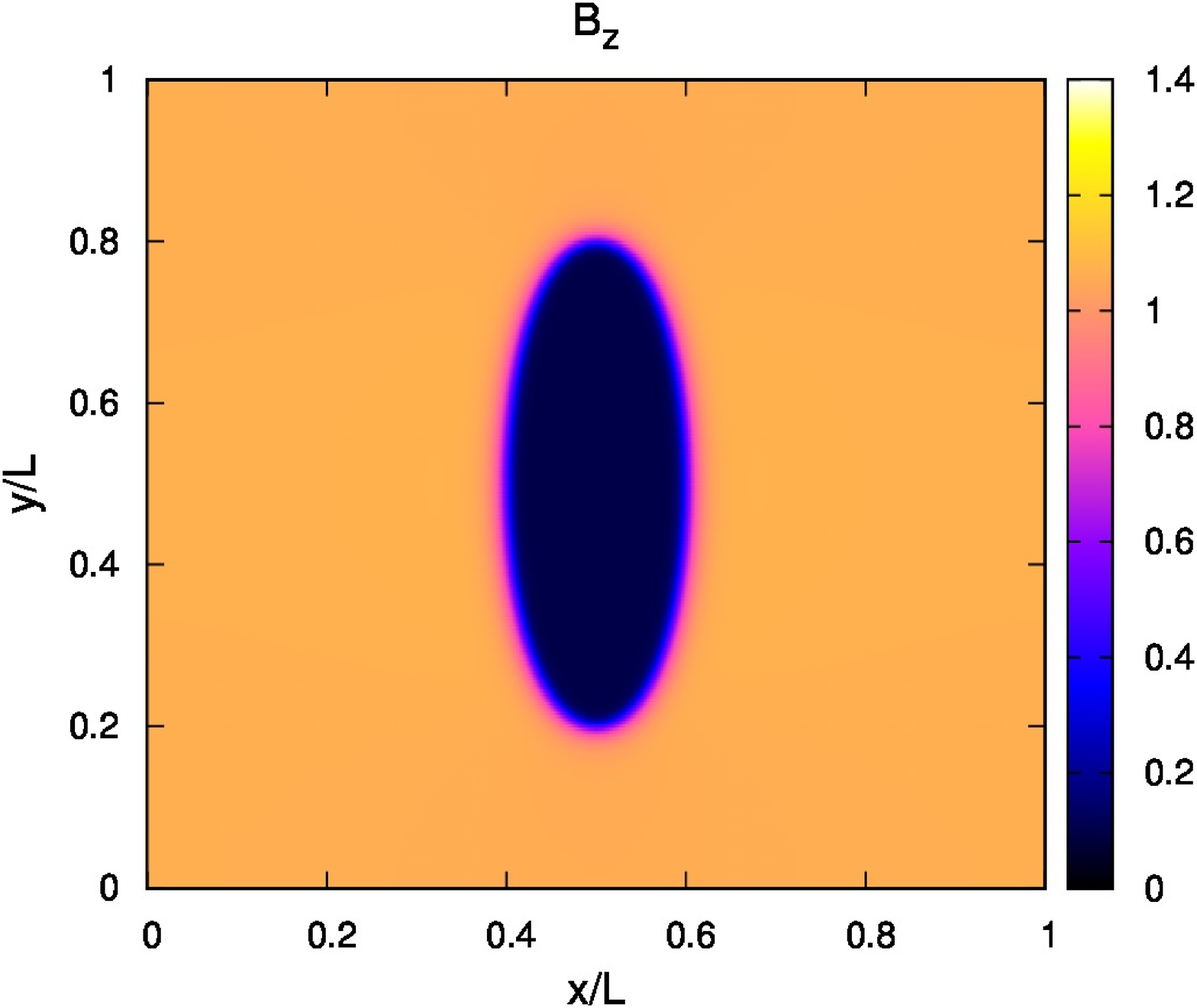,width=110mm}
\end{center}
\caption{Example of 2D ``bubble'', with $1/\protect\beta_\parallel =1.71$, $%
\langle B_y\rangle/B_0\equiv \cos\Theta_0=0.2$. }
\label{bubble}
\end{figure}

\begin{figure}[tbp]
\begin{center}
\epsfig{file=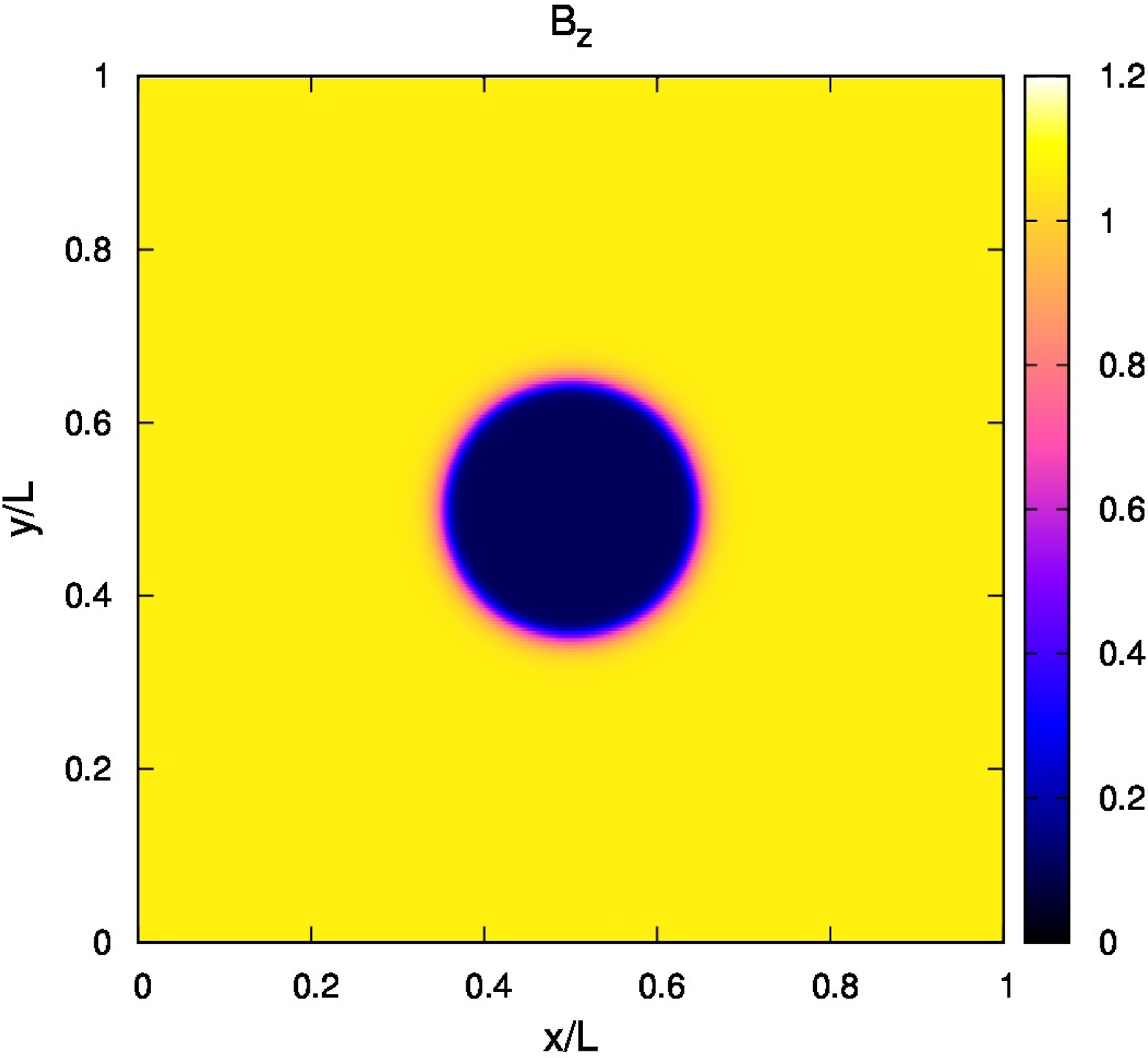,width=110mm}
\end{center}
\caption{Circular ``bubble'', with $1/\protect\beta_\parallel =1.71$, $%
\cos\Theta_0=0.0$. }
\label{bubble_circle}
\end{figure}

It is important that besides purely 1D stable configuration (``stripes''),
in our computations we have detected for some parameters also essentially 2D
stable solutions --- ``bubbles'', as shown in Fig.\ref{bubble} for $B_s=0.4$%
, $R=7.0$, $C=0.003$, $\langle B_y\rangle/B_0=0.2$, $1/\beta_\parallel =1.71$%
. In general, ``bubbles'' takes place when $B_z$ dominates, i.e. $%
\cos\Theta_0$ is sufficiently small. They have the perfect circular shape in
the case when $B_x=0$ and $B_y=0$ (see Fig.\ref{bubble_circle}). In all
cases we have inequalities $g^{^{\prime \prime }}(B_{\mathrm{in}})>0$, and $%
g^{^{\prime \prime }}(B_{\mathrm{out}})>0$, so the unstable range of $B$ is
passed in the vicinity of the bubble boundary. When $B_{\perp}=0$ the magnetic fields are constant inside and outside circle everywhere accept transient layer which is defined by the FLR.  The size of the circular patch is defined by two factors: the conservation of magnetic field flux and  the cell size. The FLR introduces small input in the this constraint, it plays a role of the surface tension.  

\begin{figure}[tbp]
\begin{center}
\epsfig{file=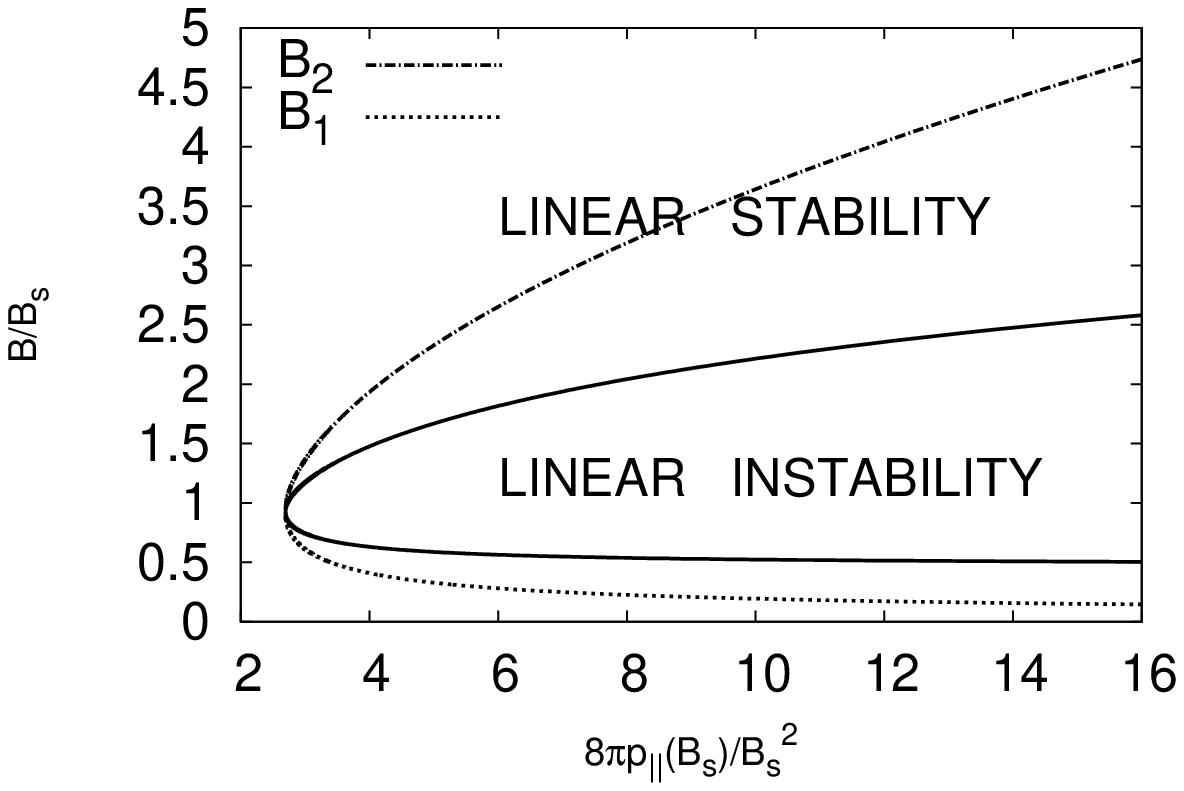,width=110mm}
\end{center}
\caption{$B_1$-$B_2$ diagram of stationary states depending on $\protect\beta%
_{\|}$ measured by $B_s$ }
\label{diagram}
\end{figure}

In Fig \ref{diagram} is shown for  circular bubbles the diagram of all possible both stable and unstable states at the fixed  $\beta_{\|}$ measured by the   
$B_s$ field. Because of the magnetic fields outside and inside the bubbles are constant, stability and instability of each state is defined by the second derivative  of the function $g(B)$. 
At the given $\beta_{\|}$ the $B_1$ and $B_2$  curves
represent the inner and outer magnetic fields when FLR is absent. The FLR in this case provides a transient solution matching the inner and outer regions. But to say that
these are the inner or outer solution one needs to have another jump, or some patch if we speak about two-dimensional structures.  Both
states $B_1$ and $B_2$ are linearly stable. These states satisfy the
necessary boundary conditions, namely, continuity of the magnetic field: $%
H_1=H_2\equiv H$, where $H$ is an additional constant. These states, thus, can
be considered as conjugated states, or, by another words, these are 
bistable states. When changing $\beta_{\|}$ which is defined  has a meaning of the parameter $\varepsilon$ we move along the curves $B_1$
and $B_2$. One should mention that in this case $\beta_{\|}$ is some
auxiliary dimensionless parameter. Real $\beta_{\|}$ is found depending on a state by means of $%
B_1$ or $B_2$. If one considers
any state $B$, say, at the given $\beta_{\|}$, without any conjugation, then
one can get linear stability or linear instability by analyzing the second
derivative sign of the function $g(B)$. The second point is that by fixing two
conjugated states one can say only that $B_2>B_1$. Only in the case when you
have another jump one can say whether it is a hole or a hump. One more point
is that the case considered here corresponds to the pure $B_z$ case when $B_{\perp}=0$.

\section{Conclusion}
In the first part of this paper we presented a review of our results concerning the weakly nonlinear regime of the mirror instability in the framework of the so-called asymptotic model. This model was demonstrated to belong to the class of the gradient systems for which the free energy can decrease in time only. In particular, it was shown that the stationary localized solutions of the model, below the mirror instability, occur unstable and, above the threshold, the system  has a blow-up behavior up to amplitude comparable with a mean magnetic field that is typical for subcritical bifurcation. We showed also that account of electrons (increase their temperature) does not change the structure of the asymptotic model. For bi-Maxwellian distribution functions for both electrons and ions all analyzed structures within the model have the form of magnetic holes. Humps can appear for distributions different from the  bi-Maxwellian ones. For instance, such situation is possible after a stage of quasi-linear relaxation ( for details see results of numerics \cite{Calif08}).  The second part of this paper contains original results concerning the possible two-dimensional mirror structures which can be formed at the saturation regime of subcritical bifurcation.   
In particular, a detailed analysis was presented for the Grad-Shafranov equations
describing static force-balanced mirror structures with anisotropic
pressures given by equations of state derived from drift kinetic equations,
when assuming an adiabatic evolution from bi-Maxwellian initial conditions.
It turns out that in two dimensions, the problem is amenable to a
variational formulation with a free energy provided by the space integral of
the parallel tension. 
Slightly below the mirror instability threshold, small
amplitude solutions associated to KPII lumps are obtained and shown to be
unstable. Based on the variational computation (the gradient method) of the stationary mirror structures,  this instability is shown to result in appearance of  one-dimensional stripes when the magnetic fields outside and inside stripes are homogeneous  with a jump which structure is defined by the FLR effects.  Such two-dimensional evolution of the stationary structures are formed for below and above threshold of the mirror instability when the $B_z$-component of magnetic field is absent. For the finite but small enough values of $B_z$ the resulting structures represent stripes. With increasing $B_z$  instead of stripes we observed in numerical simulations the formation of magnetic bubbles with the homogeneous magnetic field inside the bubbles. When $B_z$ becomes larger $B_{\perp}$ the form of bubbles change their form from elliptic to the circular one when $B_{\perp}=0$. In the latter case, the magnetic field outside and inside bubbles occurs constant and undergoes jump due to the FLR effects while crossing the bubble.  In this case, the FLR effects play the role of surface tension.  Note also, when considering stable subcritical structures, the
drift kinetic approximation breaks down, as the deep magnetic holes obtained
by a gradient method appear to be strongly sensitive to the regularization
process, an effect which in a more realistic description could be provided
by FLR corrections and/or particle trapping.

\vspace{0.5cm}
\noindent
{\bf Acknowledgments}. This work was supported
by CNRS PICS programme 6073 and RFBR grant 12-02-91062-CNRS-a.  The work of E.K. and V.R.
was also supported by the RAS Presidium Program "
Nonlinear dynamics in mathematical and physical sciences" and Grant NSh 3753.2014.2.

\end{document}